\documentclass[twocolumn,twocolappendix]{aastex631}
\usepackage{needspace}

\DeclareRobustCommand{\VAN}[3]{#2}
\let\VANthebibliography\thebibliography
\def\thebibliography{\DeclareRobustCommand{\VAN}[3]{##3}\VANthebibliography}

\newcommand{\msun}{\ensuremath{\mbox{M}_{\odot}}}
\newcommand{\lsun}{\ensuremath{\mbox{L}_{\odot}}}
\newcommand{\rsun}{\ensuremath{\mbox{R}_{\odot}}}

\newcommand{\teff}{\ensuremath{T_{\rm eff}}}

\newcommand\T{\rule{0pt}{2.6ex}}       
\newcommand\B{\rule[-1.2ex]{0pt}{0pt}} 

\newcommand{\loggp}{\ensuremath{\log(g_{\rm p})}}

\begin{document}

\title{$\epsilon$ Sagittarii: An Extreme Rapid Rotator with a Decretion Disk}

\correspondingauthor{Jeremy Bailey}
\email{j.bailey@unsw.edu.au}

\author[0000-0002-5726-7000]{Jeremy Bailey}
\affiliation{School of Physics, University of New South Wales, Sydney, NSW 2052, Australia}

\author{Fiona Lewis}
\affiliation{School of Physics, University of New South Wales, Sydney, NSW 2052, Australia}

\author[0000-0003-3476-8985]{Ian D. Howarth}
\affiliation{University College London, Gower Street, London WC1E 6BT, U.K.}

\author[0000-0003-0340-7773]{Daniel V. Cotton}
\affiliation{Monterey Institute for Research in Astronomy, 200 Eighth Street, Marina, CA, 93933, USA.}
\affiliation{Western Sydney University, Locked Bag 1797, Penrith-South DC, NSW 1797, Australia}

\author[0000-0001-6208-1801]{Jonathan P. Marshall}
\affiliation{Institute of Astronomy and Astrophysics, Academia Sinica, 11F of AS/NTU \\
Astronomy-Mathematics Building, No.1, Sec. 4, Roosevelt Rd, Taipei 106216, Taiwan}

\author[0000-0001-7212-0835]{Lucyna Kedziora-Chudczer}
\affiliation{University of Southern Queensland, Centre for Astrophysics, Toowoomba, QLD 4350, Australia}

\begin{abstract}

We report high-precision multi-wavelength linear-polarization observations of the bright B9 (or A0) star $\epsilon$~Sagittarii. The polarization shows the distinctive wavelength dependence expected for a rapidly rotating star. Analysis of the polarization data reveals an angular rotation rate $\omega$ (= $\Omega/\Omega_{\rm crit})$ of 0.995 or greater, the highest yet measured for a star in our galaxy. An additional wavelength-independent polarization component is attributed to electron scattering in a low-density edge-on gas disk that also produces the narrow absorption components seen in the spectrum. Several properties of the star (polarization due to a disk, occasional weak H$\alpha$ emission, and multiple periodicities seen in space photometry) resemble those of Be stars, but the level of activity in all cases is much lower than that of typical Be stars. The stellar properties are inconsistent with single rotating-star evolutionary tracks, indicating that it is most likely a product of binary interaction. The star is an excellent candidate for observation by interferometry, optical spectropolarimetry to detect the \"{O}hman effect, and UV polarimetry; any of which would allow its extreme rotation to be tested and its stellar properties to be refined.

\end{abstract}


\section{Introduction} \label{sec:intro}
The non-spherical shape of a rapidly rotating star results in a net linear polarization of the integrated light of the star. This effect was first predicted by \citet{harrington68}, but the resulting polarizations are small, and have only recently been observed \citep{cotton17,bailey20b,lewis22,howarth23}. The polarization has a distinctive wavelength dependence, and is strongly dependent on the rotation rate $\omega$, the axial inclination and the gravity of the star. This allows polarization data, combined with other observations, to set significant constraints on the stellar properties.

$\epsilon$~Sagittarii (HD 169022, hereafter $\epsilon$ Sgr) is a bright (V = 1.85) rapidly rotating star at a distance of 44 pc \citep{vanLeeuwen07} and with a number of anomalous properties. The spectral type is variously reported as B9$\,$IVp \citep{slettback75b}, B9.5$\,$III \citep{houk82}, A0$\,$II \citep{gray87} and A0$\,$III \citep{paunzen01}. Narrow, ``shell'' absorption components in the Balmer and other lines are seen \citep{slettback75b}, and \citet{gray87} call it a ``proto-shell star''. It has been suggested to be a $\lambda$ Boo star \citep{slettback75b}. However, \cite{baschek88} conclude, from a detailed abundance study, that $\epsilon$~Sgr, while ``peculiar in both spectrum and colors, is definitely not a $\lambda$~Boo star''. 

A companion star of magnitude V $\sim$ 7.7 at a distance of $\sim$ 2.1 arc seconds was discovered by \citet{golimowski93}. Infrared observations by \citet{hubrig01} give the K magnitude of the companion as 6.5, and are consistent with it being a main-sequence star of 0.95~\msun with \teff = 5808~K.

$\epsilon$ Sgr was found to have an infrared (IR) excess in 60 $\mu$m Infra Red Astronomical Satellite (IRAS) data by \citet{cote87}. The IRAS excess was interpreted by \citet{rhee07} as originating from a dust disk with a temperature of 100~K at a radius of 3.5 arc seconds from the star (equivalent to 150 au) and a fractional luminosity ($L_{\rm dust}/L_{\star} = 4.46\times10^{-6}$). The IR excess was also detected by Spitzer at 13 and 31 $\mu$m \citep{chen14}. 

Polarization observations of $\epsilon$~Sgr were reported by \citet{cotton16a} and \citet{bailey17}, and were larger than expected at its distance, implying the presence of intrinsic polarization. In this paper we report a more extensive set of polarization observations and compare these with models of rapidly rotating stars.

\section{Observations}

\subsection{Polarization Measurements}

\begin{table*}[t] 
\caption{Summary of Observing Runs and telescope-polarization (TP) Calibrations}
\label{tab:runs}
\begin{flushleft}
\tabcolsep 1.5 pt
\begin{tabular}{clccrcccccccrr}
\hline
\multicolumn{2}{c}{} & \multicolumn{7}{c}{Telescope and Instrument Set-Up$^a$}   &   \multicolumn{3}{c}{Observations$^b$}   &   \multicolumn{2}{c}{Calibration$^{c}$}    \\
Run & \multicolumn{1}{c}{Date Range$^d$} & Instr. &  Tel. & \multicolumn{1}{c}{f/} & Ap. & Mod. & Filt. & Det.$^e$ & $n$ & $\lambda_{\rm eff}$ &  Eff. & \multicolumn{1}{c}{$q_{\text{TP}}$} & \multicolumn{1}{c}{$u_{\rm TP}$} \\
 &  \multicolumn{1}{c}{(UT)} &  &   &  & ($\arcsec$) &  &  &  &  & (nm) & ($\%$) & \multicolumn{1}{c}{(ppm)} & \multicolumn{1}{c}{(ppm)} \\
\hline
2014AUG &  2014-09-01  &   HIPPI   & AAT  &   8  & 6.6 &  BNS-E1 &  $g^{\prime}$& B   & 1 & 463.1 & 89.0 & $-$39.9 $\pm$ 0.9 & $-$38.2 $\pm$ 0.9 \\
2016FEB/JUN$^f$ &  2016-02-27 to 06-25  &   HIPPI   & AAT  &   8  & 6.6 &  BNS-E2 &  $g^{\prime}$& B   & 2 & 463.8 & 86.9 & $-$20.5 $\pm$ 1.7 & 4.5 $\pm$ 1.8 \\
& & & & & & & $r^{\prime}$& R   & 1 & 619.9 & 80.6 & $-$8.2 $\pm$ 1.7 & 1.3 $\pm$ 1.7 \\
m2016JUN &  2016-06-10  &   M-HIPPI   & UNSW  &   11  & 58.9 &  MT &  Clear & B   & 1 & 467.9 & 75.5 & $-$64.5 $\pm$ 4.3 & $-$7.6 $\pm$ 4.4 \\
m2016JUL &  2016-08-31  &   M-HIPPI   & UNSW  &   11  & 58.9 &  MT &  Clear & B   & 1 & 464.0 & 74.3 & $-$74.5 $\pm$ 3.1 & $-$11.7 $\pm$ 3.1 \\
2017JUN & 2017-06-26 to 07-04   &   HIPPI   & AAT  &   8  & 6.6 & BNS-E2 & 425SP & B & 1 & 400.1 & 49.4 & $-$7.6 $\pm$	3.8 &	8.8 $\pm$ 3.8\\
& & & & & & & $g^{\prime}$& B   & 1 & 463.3 & 86.8 & $-$9.1 $\pm$ 1.5 & $-$2.6 $\pm$ 1.4 \\
& & & & & & & 650LP & R   & 1 & 717.6 & 62.7 & $-$8.6 $\pm$ 2.5 & $-$5.3 $\pm$ 2.5 \\
2017AUG & 2017-08-08 to 08-14   &   HIPPI   & AAT  &   8  & 6.6 & BNS-E2 & 425SP & B & 1 & 400.2 & 49.5 & $-$7.6 $\pm$	3.8 &	8.8 $\pm$ 3.8\\
& & & & & & & 500SP & B   & 1 & 435.3 & 74.0 & $-$10.0 $\pm$ 1.7 & $-$0.4 $\pm$ 1.6 \\
& & & & & & & $r^{\prime}$& R   & 2 & 619.2 & 80.7 & $-$10.6 $\pm$ 1.3 & $-$7.1 $\pm$ 1.3 \\
2018MAR & 2018-03-18 to 04-06   & HIPPI-2    & AAT  & 8*& 25.5 & BNS-E3 & $g^{\prime}$ & B   & 2 & 461.7 & 81.5 & 128.1 $\pm$ 0.9 & 3.8 $\pm$ 0.9 \\
& & & & & & & $r^{\prime}$& R   & 2 & 620.9 & 84.5 & 109.1 $\pm$ 1.4 & 6.9 $\pm$ 1.4 \\
2018JUL$^g$ & 2018-07-12 to 07-25  & HIPPI-2  & AAT  & 8*& 11.9 & BNS-E4 & $V$           & B   & 1 & 532.3 & 96.6 & $-$20.1 $\pm$ 1.5 &  2.3 $\pm$ 1.5 \\
&                         &            &      &   &      &        & $r^{\prime}$& B   & 1 & 602.0 & 88.9 & $-$10.3 $\pm$ 2.2 &  3.6 $\pm$ 2.2 \\
2018AUG$^g$ & 2018-08-21 to 09-01 & HIPPI-2   & AAT  & 8*& 11.9 & BNS-E5 & 500SP & B & 1 & 438.2 & 55.1 & 1.9 $\pm$ 1.4 & 18.0 $\pm$ 1.4 \\
& & & & & & & $V$           & B   & 1 & 532.4 & 95.7 & $-$20.1 $\pm$ 1.5 &  2.3 $\pm$ 1.5 \\
& & & & & & & 650LP & R   & 1 & 720.9 & 76.8 & $-$6.5 $\pm$ 1.8 & 3.8 $\pm$ 1.8 \\
2023APR/MAY$^h$ & 2023-06-03 & HIPPI-2   & AAT  & 15& 12.7 & ML-E1 & 425SP & B & 1 & 398.8 & 79.3 & $-$1.0 $\pm$ 2.4 & 7.8 $\pm$ 2.1 \\
& & & & & & & 500SP & B & 1 & 438.2 & 86.9 & $-$20.1 $\pm$ 1.7 & $-$5.1 $\pm$ 1.5 \\
& & & & & & & $g^{\prime}$ & B   & 1 & 460.2 & 89.2 & 5.9 $\pm$ 1.0 &  0.0 $\pm$ 1.0 \\
& & & & & & & $r^{\prime}$ & B   & 1 & 601.8 & 64.2 & 4.9 $\pm$ 4.5 &  $-$19.7 $\pm$ 3.9 \\
& & & & & & & $r^{\prime}$ & R   & 1 & 620.8 & 60.5 & $-$2.7 $\pm$ 3.2 & $-$18.8 $\pm$ 3.3 \\
& & & & & & & 650LP & R   & 1 & 719.1 & 44.6 & $-$36.6 $\pm$ 5.8 & $-$35.4 $\pm$ 6.2 \\

\hline
\end{tabular}
Notes: \\
\textbf{*} Indicates use of a 2$\times$ negative achromatic lens, effectively making the foci f/16.\\ 
\textbf{$^a$} A full description, along with transmission curves for all the components and modulation characterisation of each modulator (`Mod.') in the specified performance era, can be found in \citet{bailey20a}.\\ 
\textbf{$^b$} Mean values are given as representative of the observations made of $\epsilon$~Sgr. Individual values are given in Table~\ref{tab:observations} for each observation; $n$ is the number of observations of $\epsilon$~Sgr.\\
\textbf{$^c$} Most of the observations used to determine the TP and the high-polarization standards observed to calibrate position angle (PA), are described in \citet{bailey17} (m2016JUN, m2016JUL), \citet{bailey15} (2014MAY), \citet{cotton17b} (2016FEB, 2016JUN) \citet{cotton19b} (2017JUN, 2017AUG), and \citet{bailey20a} (other runs). This is the first data presented from the 2023MAY run; 2-6 total observations per band of HD 2151, HD 61421, HD 102647, HD 102870 or HD 140573 were used to determine TP; PA was calibrated with reference to HD 84810, HD 161471, HD 111613 and HD 183143.\\   
\textbf{$^d$} Dates given are for observations of $\epsilon$~Sgr and/or control stars. \\
\textbf{$^e$} B, R indicate blue- and red-sensitive H10720-210 and H10720-20 photomultiplier-tube detectors, respectively.\\
\textbf{$^f$} TP calibration was carried out using observations combined from 2016FEB and 2016JUN runs. \\
\textbf{$^g$} TP calibration was carried out using observations combined from 2018JUL and 2018AUG runs. \\
\textbf{$^h$} TP calibration was carried out using observations combined from 2023APR\,A/B and 2023MAY runs. \\
\end{flushleft}
\label{tab:mod}
\end{table*}

\begin{table*}[t] 
\caption{Polarization Observations of $\epsilon$~Sgr (Sorted by Effective Wavelength).}
\begin{flushleft}
\label{tab:observations}
\tabcolsep 2 pt
\begin{tabular}{llrrccrrrrrr}
\hline
\multicolumn{1}{c}{Run}     & \multicolumn{1}{c}{UT (mid-point)}                  & Dwell & Exp. & Filt. & Det.$^a$ & $\lambda_{\rm eff}$ & Eff. & \multicolumn{1}{c}{$q$} & \multicolumn{1}{c}{$u$} & \multicolumn{1}{c}{$p$} & \multicolumn{1}{c}{$\theta$}\\
        &                       & \multicolumn{1}{c}{(s)} & \multicolumn{1}{c}{(s)} & & & \multicolumn{1}{c}{(nm)} & \multicolumn{1}{c}{(\%)} & \multicolumn{1}{c}{(ppm)} & \multicolumn{1}{c}{(ppm)} & \multicolumn{1}{c}{(ppm)} & \multicolumn{1}{c}{($^\circ$)}\\
\hline
2023MAY & 2023-06-03 16:39:20 & 1615 & 1280 & 425SP & B & 398.8 & 79.3 & $-$5.8 $\pm$ 12.1 & $-$13.6 $\pm$ 12.0 & 14.8 $\pm$ 12.0 & 123.6 $\pm$ 27.9 \\
2017JUN & 2017-06-26 16:29:55 & 1557 & 800 & 425SP & B & 400.1 & 49.4 &  6.7 $\pm$ 15.6 &   $-$50.3 $\pm$ 15.6 & 50.7 $\pm$ 15.6 &   138.8 $\pm$ \phantom{0}9.2 \\
2017AUG & 2017-08-10 14:14:07 & 2124 & 800 & 425SP & B & 400.2 & 49.5 &  28.1 $\pm$ 14.6 &   $-$19.4 $\pm$ 14.5 & 34.1 $\pm$ 14.6 &  162.7 $\pm$ 14.2\\
2023MAY & 2023-06-03 17:03:20 & 996 & 640 & 500SP & B & 433.4 & 86.9 & 36.0 $\pm$ \phantom{0}6.7 & 60.4 $\pm$ \phantom{0}6.7 & 70.3 $\pm$ \phantom{0}6.7 & 29.6 $\pm$ \phantom{0}2.7 \\
2017AUG & 2017-08-11 13:53:51 & 1506 & 640 & 500SP & B & 435.3 & 74.0 & 21.8 $\pm$ \phantom{0}8.9 &  81.0 $\pm$ \phantom{0}8.9 &  83.9 $\pm$  \phantom{0}8.9 &  37.5 $\pm$   \phantom{0}3.0\\
2018AUG & 2018-08-21 14:16:31 & 973 & 640 & 500SP & B & 438.2 & 54.9 & 25.1 $\pm$ \phantom{0}6.8 &  83.2 $\pm$ \phantom{0}6.8 &   86.9 $\pm$   \phantom{0}6.8 &   36.6 $\pm$   \phantom{0}2.2\\
2023MAY & 2023-06-03 16:17:39 & 983 & 640 & $g^{\prime}$ & B & 460.2 & 89.2 & 19.3 $\pm$ \phantom{0}2.5 & 95.1 $\pm$ \phantom{0}2.6 & 97.0 $\pm$ \phantom{0}2.6 & 39.3 $\pm$ \phantom{0}0.8 \\
2018MAR$^s$ & 2018-04-06 17:39:10 & 1032 & 640 & $g^{\prime}$ & B & 461.7 & 81.5 & 71.3 $\pm$ \phantom{0}2.5 & 99.9 $\pm$ \phantom{0}2.5  &  122.7 $\pm$   \phantom{0}2.6 &   27.2 $\pm$   \phantom{0}0.6\\
2018MAR$^l$ & 2018-04-06 17:19:21 & 1189 & 640 & $g^{\prime}$ & B & 461.8 & 81.6 & 16.2 $\pm$ \phantom{0}2.9 & 95.5 $\pm$ \phantom{0}2.7  &  96.9 $\pm$   \phantom{0}2.8 &   40.2 $\pm$   \phantom{0}0.8\\
2014AUG & 2014-09-01 10:11:12 & 1843 & 640 & $g^{\prime}$ & B & 463.1 & 89.0 & 41.0 $\pm$ \phantom{0}3.6 &  164.0 $\pm$ \phantom{0}3.6 &  169.0 $\pm$   \phantom{0}3.6 &   38.0 $\pm$   \phantom{0}0.6 \\
2017JUN & 2017-06-26 16:02:50 & 1588 & 800 & $g^{\prime}$ & B & 463.3 & 86.8 & 13.4 $\pm$ \phantom{0}6.0 &   102.0 $\pm$ \phantom{0}4.2 &  102.9 $\pm$   \phantom{0}5.1 &  41.3 $\pm$   \phantom{0}1.7\\
2016FEB & 2016-03-01 18:37:33 & 973 & 480 & $g^{\prime}$ & B & 463.8 & 86.9 & 27.6 $\pm$ \phantom{0}3.7 &   88.8 $\pm$ \phantom{0}3.7 &  93.0 $\pm$   \phantom{0}3.7 &  36.4 $\pm$   \phantom{0}1.2\\
2016FEB & 2016-02-29 18:34:56 & 1012 & 480 & $g^{\prime}$ & B & 463.9 & 87.0 & 36.2 $\pm$ \phantom{0}4.0 &   85.1 $\pm$ \phantom{0}4.1 &  92.5 $\pm$   \phantom{0}4.0 &  33.5 $\pm$   \phantom{0}1.2\\
m2016JUL & 2016-08-31 10:37:32 & 1609 & 800 & Clear & B & 464.0 & 74.3 & 43.5 $\pm$ 19.9 &   123.2 $\pm$ 20.6 &  130.7 $\pm$   20.3 &  35.3 $\pm$   \phantom{0}4.4\\
m2016JUN & 2016-06-10 10:47:32 & 1850 & 800 & Clear & B & 467.9 & 75.5 & 29.5 $\pm$ 21.9 &   98.45 $\pm$ 21.8 &  102.7 $\pm$   21.8 &  36.7 $\pm$   \phantom{0}6.2\\
2018JUL & 2018-07-25 16:08:13 & 1025 & 640 & $V$ & B & 532.3 & 96.6 & 39.8 $\pm$ \phantom{0}4.1 &  217.1 $\pm$ \phantom{0}4.1 &  220.7 $\pm$   \phantom{0}4.1 &  39.8 $\pm$   \phantom{0}0.5\\
2018AUG & 2018-08-21 14:36:42 &1345 & 960 & $V$ & B & 532.4 & 95.7 &  45.2 $\pm$ \phantom{0}3.1 &  178.5 $\pm$ \phantom{0}3.0 &  184.1 $\pm$   \phantom{0}3.1 &   37.9 $\pm$   \phantom{0}0.5\\
2023MAY & 2023-06-03 17:23:21 & 1405 & 960 & $r^{\prime}$ & B & 601.8 & 64.2 & 31.5 $\pm$ \phantom{0}7.5 &  261.8 $\pm$ \phantom{0}6.7 & 263.7 $\pm$ \phantom{0}7.1 & 41.6 $\pm$ \phantom{0}0.8 \\ 
2018JUL & 2018-07-25 16:26:18 & 1045 & 710 & $r^{\prime}$ & B & 602.0 & 88.9 & 51.0 $\pm$ \phantom{0}7.3 &  262.8 $\pm$ \phantom{0}6.4 & 267.7 $\pm$   \phantom{0}6.8 &   39.5 $\pm$   \phantom{0}0.8\\
2017AUG & 2017-08-09 13:12:24 & 1880 & 640 & $r^{\prime}$ & R & 619.2 & 80.7 & 61.1 $\pm$ \phantom{0}3.9 &  281.3 $\pm$ \phantom{0}3.7 &  287.9 $\pm$   \phantom{0}3.8 &  38.9 $\pm$   \phantom{0}0.4\\
2017AUG & 2017-08-08 14:18:24 & 1339 & 640 & $r^{\prime}$ & R & 619.3 & 80.7 & 63.8 $\pm$ \phantom{0}3.7 &  283.9 $\pm$ \phantom{0}3.7 &  291.0 $\pm$   \phantom{0}3.7 &  38.7 $\pm$   \phantom{0}0.4\\
2016FEB & 2016-02-27 16:41:49 & 1237 & 640 & $r^{\prime}$ & R & 619.9 & 80.6 & 30.3 $\pm$ \phantom{0}5.2 &  227.5 $\pm$ \phantom{0}5.4 &  229.5 $\pm$   \phantom{0}5.3 &  41.2 $\pm$   \phantom{0}0.7\\
2023MAY & 2023-06-03 18:24:03 & 958 & 640 & $r^{\prime}$ & R & 620.8 & 60.5 & 64.5 $\pm$ \phantom{0}5.1 & 284.8 $\pm$ \phantom{0}5.2 & 292.0 $\pm$ \phantom{0}5.2 & 38.6 $\pm$ \phantom{0}0.5 \\
2018MAR$^l$ & 2018-03-18 18:49:50 & 980 & 640 & $r^{\prime}$ & R & 620.9 & 84.5 & 78.4 $\pm$ \phantom{0}3.2 &  276.1 $\pm$ \phantom{0}3.2 &  287.0 $\pm$   \phantom{0}3.2 &  37.1 $\pm$   \phantom{0}0.3\\
2018MAR$^s$ & 2018-03-18 19:11:58 & 1300 & 640 & $r^{\prime}$ & R & 620.9 & 84.5 & 71.9 $\pm$ \phantom{0}4.6 &  278.7 $\pm$ \phantom{0}4.4 &  287.8 $\pm$   \phantom{0}4.4 &  37.8 $\pm$   \phantom{0}0.5\\
2017JUN & 2017-07-04 13:08:09 & 2165 & 1440 & 650LP & R & 717.6 & 62.7 & 81.9 $\pm$ \phantom{0}4.8 &  328.5 $\pm$ \phantom{0}4.8  &  338.6 $\pm$   \phantom{0}4.8 &   38.0 $\pm$   \phantom{0}0.4\\
2023MAY & 2023-06-03 18:45:19 & 1634 & 1280 & 650LP & R & 719.1 & 44.6 & 101.3 $\pm$ \phantom{0}8.1 & 341.2 $\pm$ \phantom{0}8.5 & 355.9 $\pm$ \phantom{0}8.3 & 36.7 $\pm$ \phantom{0}0.7 \\
2018AUG & 2018-08-23 11:47:56 & 1232 & 960 & 650LP & R & 720.9 & 76.8 & 52.4 $\pm$ \phantom{0}4.1 &  318.2 $\pm$ \phantom{0}4.1 &  322.5 $\pm$   \phantom{0}4.1 &   40.3 $\pm$   \phantom{0}0.4\\
\hline
\end{tabular}
{$^a$} B, R indicate blue- and red-sensitive H10720-210 and H10720-20 photomultiplier-tube detectors, respectively.
{$^{s/l}$} Aperture sizes are as given in Table \ref{tab:mod} except where indicated by these superscripts; $l$: 25.5 arc seconds, $s$: 5.3 arc seconds -- small enough to exclude the reported debris disk but not the binary companion \citep{rodriguez12}.   
\end{flushleft}
\end{table*}

We report here 28 linear polarization observations of $\epsilon$~Sgr. Of these, 26 were made on the 3.9-m Anglo-Australian Telescope (AAT) at Siding Spring Observatory using either the HIPPI \citep[high precision polarimetric instrument,][]{bailey15} or HIPPI-2 \citep{bailey20a} polarimeters. Two observations were made with the Mini-HIPPI \citep[]{bailey17} polarimeter on the 35-cm telescope at the University of New South Wales (UNSW) observatory in Sydney. These 3 polarimeters all use similar techniques, making use of ferro-electric liquid-crystal modulators operating at 500~Hz, and compact photomultiplier tubes (PMTs) as detectors. Some of the observations were previously reported by \citet{cotton16a} and \citet{bailey17}, but all the data presented here have been re-analysed using the methods described by \citet{bailey20a}. We adopt a new calibration of modulator performance that has been carried out for an upcoming paper \citep{cotton24}. The resulting final polarization values are similar to past work.

Details of the observing runs, made between 2014 and 2023, as well as instrumental details are listed in Table \ref{tab:mod}. The polarization observations of $\epsilon$ Sgr are listed in Table~\ref{tab:observations}. New observations of inter\-stellar control stars are described in Appendix \ref{apx:interstellar}. The AAT observations were obtained with a range of broad-band filters. The Sloan Digital Sky Survey (SDSS) $g^\prime$ and $r^\prime$ filters used with HIPPI were made by Omega Optics. For HIPPI-2 the corresponding filters were generation-2 filters from Astrodon Photometrics. The two Mini-HIPPI observations were made with no filter (Clear). A bandpass model as described in \citet{bailey20a} was used to determine the effective wavelength and modulation efficiency for each observation. This model takes account of all the optical components as well as the source spectrum, and atmospheric transmission. These values are listed in Table \ref{tab:observations} and the effective wavelengths for these observations range from $\sim$400~nm to $\sim$720~nm.

\subsection{Imaging}

\label{sec:imaging}

Infrared images of $\epsilon$~Sgr were obtained on 2016 Sep 10 using the IRIS2 instrument \citep{tinney04} on the 3.9m Anglo-Australian Telescope in a narrow band filter at a wavelength of 2.3~$\mu$m. The separation ($\rho$) and position angle of the companion were measured from the images with the Python Photutils package \citep{bradley22} using profile fitting methods based on the DAOPHOT algorithms \citep{stetson87}. The results, together with those from past observations, are listed in Table \ref{tab:binary}. The separation of the companion has been slowly increasing, with little change to the position \mbox{angle}. 

The companion was included in the aperture for all our polarization observations. However, it contributes less than 1\% of the optical-region light, and as a main-sequence G star is not expected to have significant intrinsic polarization \citep{cotton17b}.

\begin{table}
    \centering
    \caption{Measurements of the $\epsilon$~Sgr Binary System.}
    \begin{flushleft}
    \begin{tabular}{lllll}
    \hline  UT Date & $\rho$ (\arcsec) & pa (\degr) & $\Delta$m (mag) & Ref \\ \hline
    1992-06-18 & 2.06$\pm$0.01 & 142$\pm$1 & 5.85 (V) & 1  \\
    1999-03$^a$       & 2.39    &  142    & 4.6 (K) & 2 \\
    2016-09-10 & 2.58$\pm$0.04 & 140$\pm$1 & 4.2 (2.3$\mu$m) & 3 \\
    \hline
    \end{tabular}
    Notes: \\
    References: 1 \cite{golimowski93}, 2 \cite{hubrig01}, 3. This paper (Section \ref{sec:imaging}) \\
    $^a$ The exact date of the observation is not reported. \\
    \end{flushleft}
\label{tab:binary}
\end{table}

\subsection{Archival Spectroscopy}

\label{sec:archival_spect}

Spectra of $\epsilon$~Sgr have been obtained from the archive of the Canada France Hawaii Telescope (CFHT). These were taken with the Echelle Spectropolarimetric Device for Observation of Stars (ESPaDOnS) instrument \citep{donati03} on 2014 Sep 8. The spectra cover 370 to 1000 nm at R $\sim$ 70000. The individual integrations have been combined to form a single spectrum which was used for the analysis in Section \ref{sec:variability} and \ref{sec:grid}.

We also used three spectra covering the H$\alpha$ region taken with the ultraviolet and visual echelle spectrograph \citep[UVES,][]{dekker00} on the European Southern Observatory (ESO) Very Large Telescope (VLT)\footnote{Based on data obtained from the ESO Science Archive Facility with DOI: \dataset[10.18727/archive/50]{\doi{10.18727/archive/50}}}. These cover 472.6 to 683.5 nm and have R = 74450. The dates of observation were 2016 Apr 22, May 17 and Jun 1.

\section{Modelling}

\subsection{Polarization of Rotating Stars}

\label{sec:polmod}

We model the polarization of rotating stars using the methods described in detail by \cite{bailey20b} and also used by \cite{cotton17}, \cite{lewis22} and \cite{howarth23}. We use a Roche model for the rotating star and assume gravity darkening according to \cite{espinosa11}. This allows us to determine the distribution of local temperature and gravity as a function of latitude. We use a set of custom ATLAS9 \citep{castelli03} models to represent the local stellar atmosphere at 46 colatitudes from 0 to 90\degr\ at 2\degr\ steps. For each of these model atmospheres we then calculate the specific intensity and polarization of the emergent radiation. These calculations use a version of the SYNSPEC spectral synthesis code \citep{hubeny85,hubeny12} which we have modified to do full polarized radiative transfer using the vector linearized discrete ordinate radiative transfer (VLIDORT) code of \citet{spurr06}.

We overlay a rectangular grid of pixels over the projected view of the star. For each pixel we determine the local temperature, gravity, and the surface normal viewing angle. We then use these values to interpolate in our set of radiative transfer calculations to determine the local specific intensity and polarization. Summing over all the pixels then gives the integrated flux and polarization as a function of wavelength.

\begin{figure}
    \centering
    \includegraphics[width=\columnwidth]{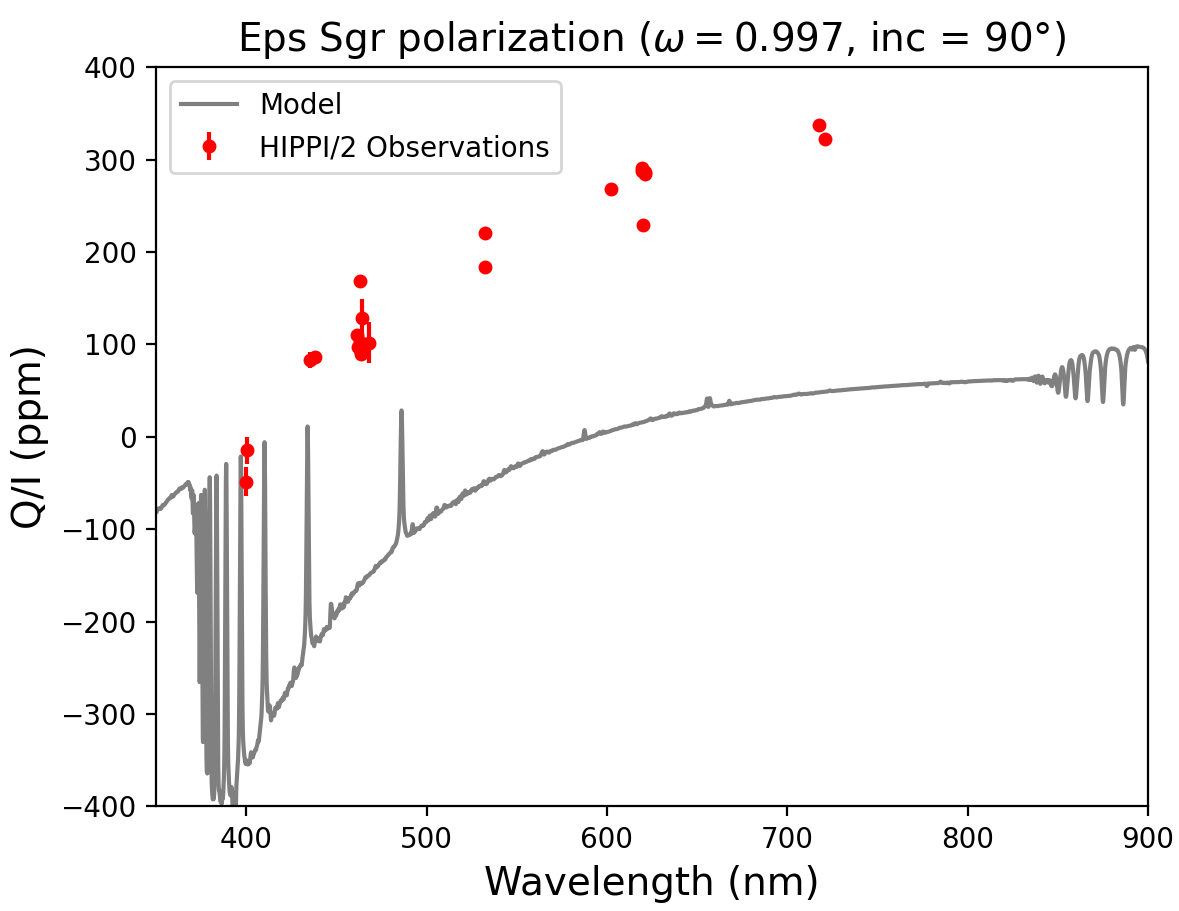}
    \caption{Polarization observations (rotated to put all the polarization into the Q stokes parameter) compared with a model of polarization for a rapidly rotating star. The observations show similar wavelength dependence to the model but the polarization is offset by $\sim$300 ppm in the positive direction.}
    \label{fig:obs_model}
\end{figure}

\needspace{2\baselineskip}
\subsection{Polarization Components} 
\label{sec:components}

Figure \ref{fig:obs_model} shows the polarization observations rotated through $\sim$40\degr, which puts the polarization primarily in the Q Stokes parameter. The polarization shows a large change with wavelength from values of about -50 ppm at $\sim$400 nm, to more than 300 ppm at $\sim$720 nm. This large wavelength change matches that expected for a star with a large rotation rate. However, when we compare the observations with a rapidly-rotating-star model, as shown in Figure \ref{fig:obs_model}, we find that the observed polarization is offset in a positive direction from the modelled value by $\sim$300 ppm. In this plot negative polarization is perpendicular to the star's rotation axis, positive polarization is parallel to the axis.

For the polarization to be consistent with rotational distortion, there must be an additional source of polarization of $\sim$300 ppm with a position angle parallel to the star's rotation axis. In our previous studies of rotating stars we have been able to model the observations using a rotating star model and a small contribution of interstellar polarization. For $\epsilon$ Sgr at a distance of 44 pc \citep{vanLeeuwen07} the expected level of interstellar polarization is $\sim$60 ppm according to the models of \citet{cotton17b} and a value as high as 300 ppm would be extremely unlikely (see Appendix \ref{apx:interstellar}). Furthermore the interstellar polarization would have to be aligned with the rotation axis of the star to provide the required offset. This would not be consistent with what is measured for nearby stars and seems intrinsically unlikely. 

For $\epsilon$ Sgr there is, however, an additional likely source of polarization with the required properties. The narrow shell absorption features in the spectrum, as described in Section \ref{sec:intro}, indicates the presence of circumstellar gas. In such a rapidly rotating star this is most likely in the form of a narrow equatorial disk. Such disks are common in rapidly rotating B stars as seen in the classical Be stars \citep{rivinius13}. One of the important observable features of Be stars is polarization due to starlight scattered from the disk. The typical polarization level in nearby Be stars (where inter\-stellar polarization is unimportant) is $\sim$3000--6000 ppm \citep{cotton16a}.

The much lower polarization ($\sim$300 ppm) we infer for the disk in $\epsilon$ Sgr, as well as the lack of strong H$\alpha$ emission, indicates a disk of much lower density than those usually found in classical Be stars. At low densities we expect electron scattering to be the dominant opacity source leading to a flat, wavelength-independent polarization \citep{halonen13,rivinius13}. This differs from the wavelength-dependent polarization structure seen in the higher-density disks of classical Be stars \citep*{poeckert79,wood97} where bound--free absorption is important. Comparison with the models of \citet{halonen13} suggests a gas density of $\sim10^{-13}$ g cm$^{-3}$ whereas densities up to $\sim10^{-10}$ g cm$^{-3}$ are seen in classical Be stars.

The polarization position angle of a Be star disk is perpendicular to the disk \citep{quirrenbach97}. Hence the polarization is parallel to the star's rotation axis, as required for our offsetting polarization component.

\subsection{Polarization Variability}
\label{sec:variability}

The polarization observations of $\epsilon$ Sgr (Table \ref{tab:observations}) show considerable scatter between observations in the same filter on different dates. The total range in $p$ values is 77 ppm at $g^\prime$ and 62 ppm at $r^\prime$. This is much larger than the statistical errors, and much larger than the scatter seen in our similar observations of other rapidly rotating stars using the same instruments \citep[e.g.][]{bailey20b,howarth23}. The most extreme values are seen in some of the earliest observations in 2014 and 2016, whereas observations closely spaced in time generally show good agreement. This suggests variability of the polarization on long timescales. Such variability is not unexpected for a system with a disk component as described in the previous section. Polarization in Be stars is often variable \citep{coyne76}. 

\begin{figure}
    \centering
    \includegraphics[width=\columnwidth]{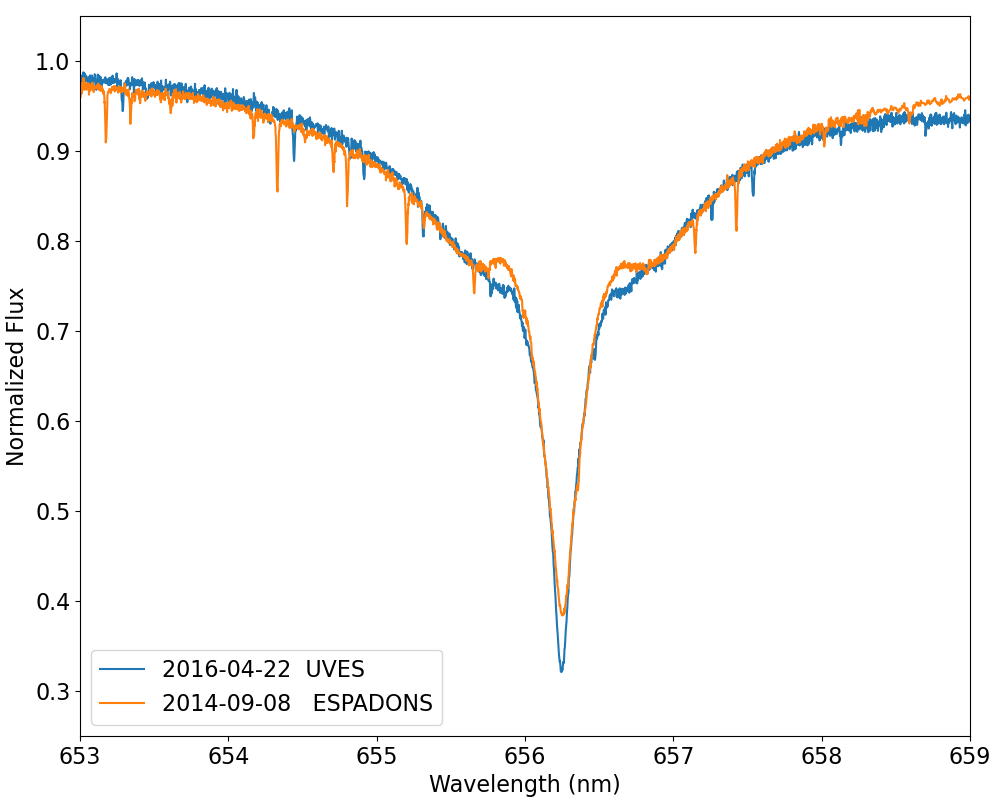}
    \caption{Spectra of $\epsilon$ Sgr in the H$\alpha$ region. The UVES (ESO VLT) spectrum is from 2016 Apr 22 and shows broad absorption and the narrow shell absorption component. The ESPaDOnS (CFHT) spectrum on 2014 Sep 8 shows additionally double-peaked core emission. The very narrow absorption features are telluric lines.}
    \label{fig:uves_cfht}
\end{figure}

Independent evidence for a variable Be-type disk in $\epsilon$ Sgr comes from the archival spectroscopy (Section \ref{sec:archival_spect}). Figure~\ref{fig:uves_cfht} shows two spectra of $\epsilon$ Sgr. One of these, the CFHT spectrum from 2014 Sep 8, shows weak double-peaked emission in H$\alpha$. The other spectrum, taken on 2016 Apr 22 with VLT UVES shows a broad photospheric absorption and narrow shell-absorption core. Similar behavior is seen in the other UVES spectra from 2016. Since the earlier spectroscopy discussed in Section \ref{sec:intro} makes no mention of emission we assume the 2016 spectra represent the more usual state. Interestingly the highest $g^\prime$ polarization was observed on 2014 Sep 1, only a few days before the spectrum showing the emission, whereas polarization observations in 2016 show some of the lowest values.  

\needspace{2\baselineskip}
\subsection{Model Grid}

\label{sec:grid}

\begin{figure}
    \centering
    \includegraphics[width=\columnwidth]{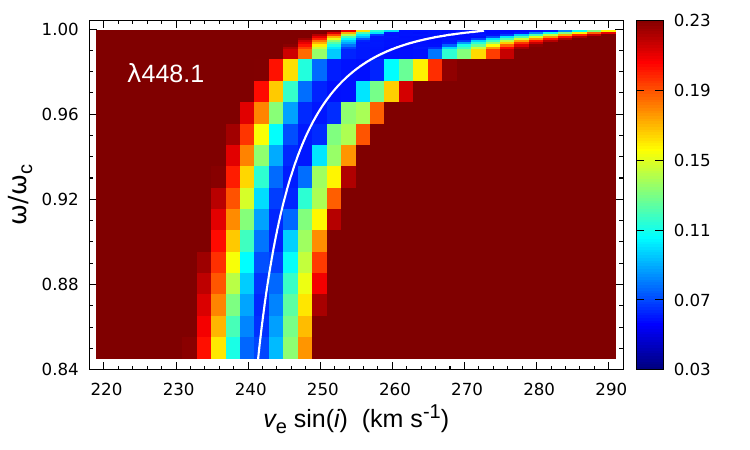}
    \caption{Results of Fourier-Transform modelling of CFHT spectra of the MgII 448.1nm line to determine $v_{\rm e}\sin{i}$. The value plotted is the rms difference between observed and modelled Fourier Transforms.}
    \label{fig:vsini}
\end{figure}
The polarization of a rotating-star model (as described in Section \ref{sec:polmod}) depends on four main parameters: the polar temperature and gravity ($T_p$, $g_p$), the rotation rate ($\omega$ = $\Omega/\Omega_{\rm crit}$){\footnote{Where the subscript `crit' is used to denote critical rotation throughout this paper.}, and the inclination of the star's rotation axis ($i$). The polarization observations alone are not sufficient to determine these four parameters so, as in our previous studies \citep{bailey20b,lewis22,howarth23} we use additional information to provide relationships between these parameters.

The first of these is the spectral-energy distribution, which 
principally constrains the temperature and angular diameter (i.e., radius).
We use photometry from \citet{johnson66} (optical) and archival spectroscopy obtained with the International Ultraviolet Explorer (IUE; UV). IUE spectra taken in low-resolution mode
($\Delta\lambda \simeq 0.6$~nm) through the spectrographs' large aperture provide the most reliable spectro\-photometry.  Only one spectrum is available for each of the short- and long-wavelength ranges ($\lambda\lambda \simeq 115$--198, 185--335~nm), but the data are well exposed  (SWP~16443, LWR~12689). We adjusted the fluxes to match the current HST calibration by using results from \citet{massa20}, including the minor update described by \citet{fitzp19}. This gives an $\sim$8\%\ increase in integrated UV flux over the standard pipeline product (and a $\sim$1.5\%\ increase in the inferred $\teff$). The resulting integrated observed flux in the adopted 123.5--321~nm region is 1.40 $\times$ 10$^{-6}$ erg cm$^{-2}$ s$^{-1}$. We then applied an extinction correction for
$E(B-V) = 0\fm008$, estimated using the \textsc{g-tomo} tool\footnote{\texttt{https://explore-platform.eu/sdas}}  (based on results from \citealt{lallement22}), with a \citet{seaton79} extinction curve (adopting $R \equiv A(V)/E(B-V) = 3.1$).

The second constraint is the value of $v_{\rm e}\sin{i}$. We determined new values of $v_{\rm e}\sin{i}$ using a Fourier Transform analysis of archival CFHT spectra. We used the MgII 448.1~nm line, chosen as the rotational line shape is less affected by shell absorption than alternatives such as SiII 634.7~nm. The analysis was done by comparing model spectra, calculated using Roche geometry, to the observations in Fourier-Transform space \citep[see][for a more detailed description of this approach]{howarth23}. Because of the rotational distortion of the star and the strong gravity darkening the inferred $v_{\rm e}\sin{i}$ values depend on the rotation rate $\omega$, as shown in Figure~\ref{fig:vsini}. For slow rotations (alone) the inferred $v_{\rm e}\sin{i}$ is in reasonable agreement with the value of  236~km~s$^{-1}$ given by \citet{royer02a}.

We adopt the $v_{\rm e}\sin{i}\text{--}\omega$ relation of Figure~\ref{fig:vsini}, together with the the Hipparcos parallax of 22.76$\pm$0.24 mas (\citealt{vanLeeuwen07}; the star is too bright for Gaia).  We can then construct a two-dimensional grid of stellar parameters where all other values can be established from the two parameters $\omega$ and $i$. Each model in this grid corresponds to a star that fits the spectral-energy distribution and $v_{\rm e}\sin{i}$. We can then calculate the polarization for each of these grid models and compare with the observations.

\begin{figure}
    \centering
    \includegraphics[width=\columnwidth]{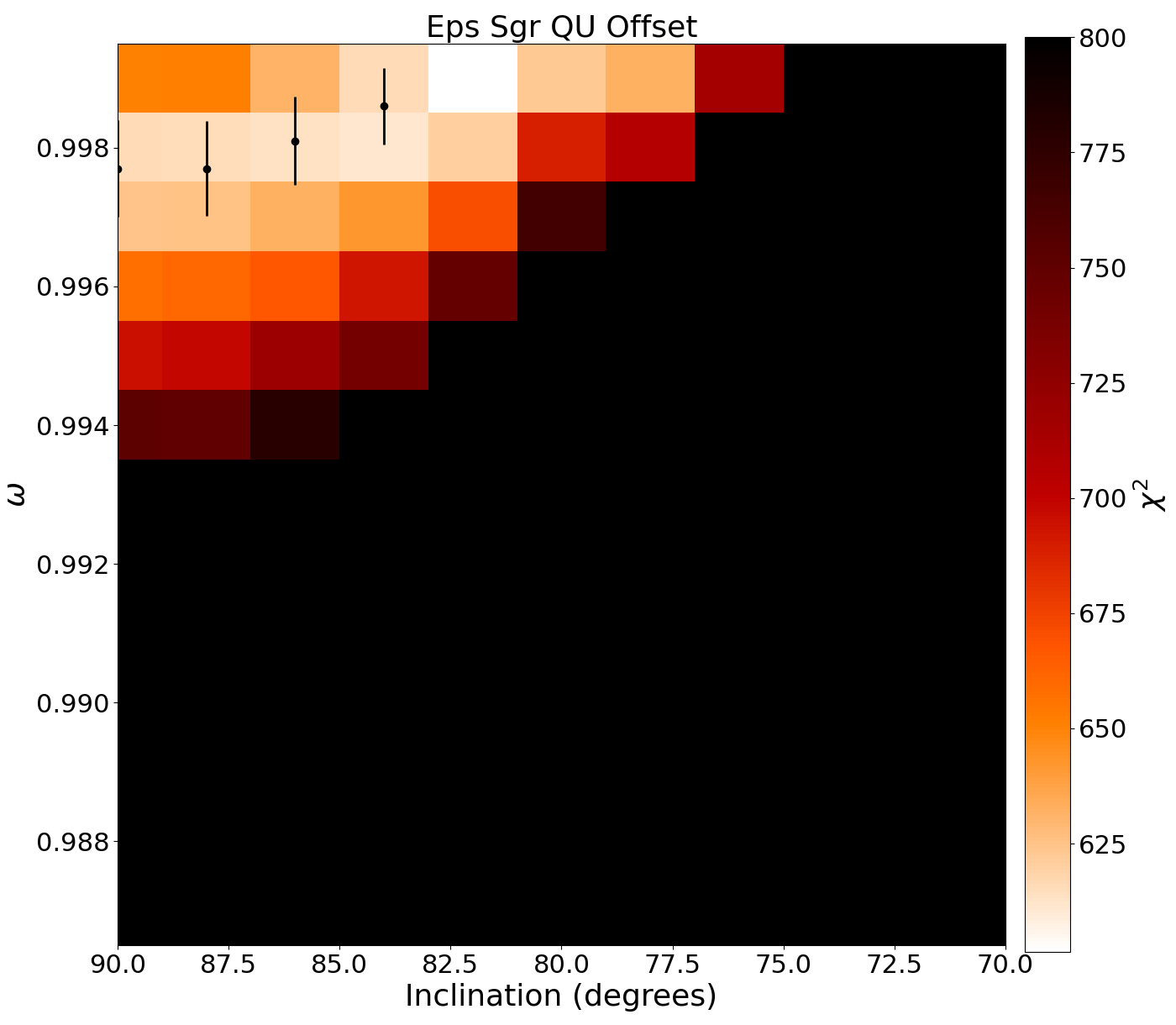}
    \caption{Fit to $\epsilon$~Sgr grid using the rotational polarization models plus  a fixed polarization offset. In this and similar figures each point on the grid corresponds to a set of stellar parameters that fits the spectral energy distribution and $v_{\rm e}\sin{i}$ of $\epsilon$ Sgr as described in Section \ref{sec:grid}. The plot shows how well each of these grid models fits the polarization observations.}
    \label{fig:fit_offset1}
\end{figure}

\begin{figure}
    \centering
    \includegraphics[width=\columnwidth]{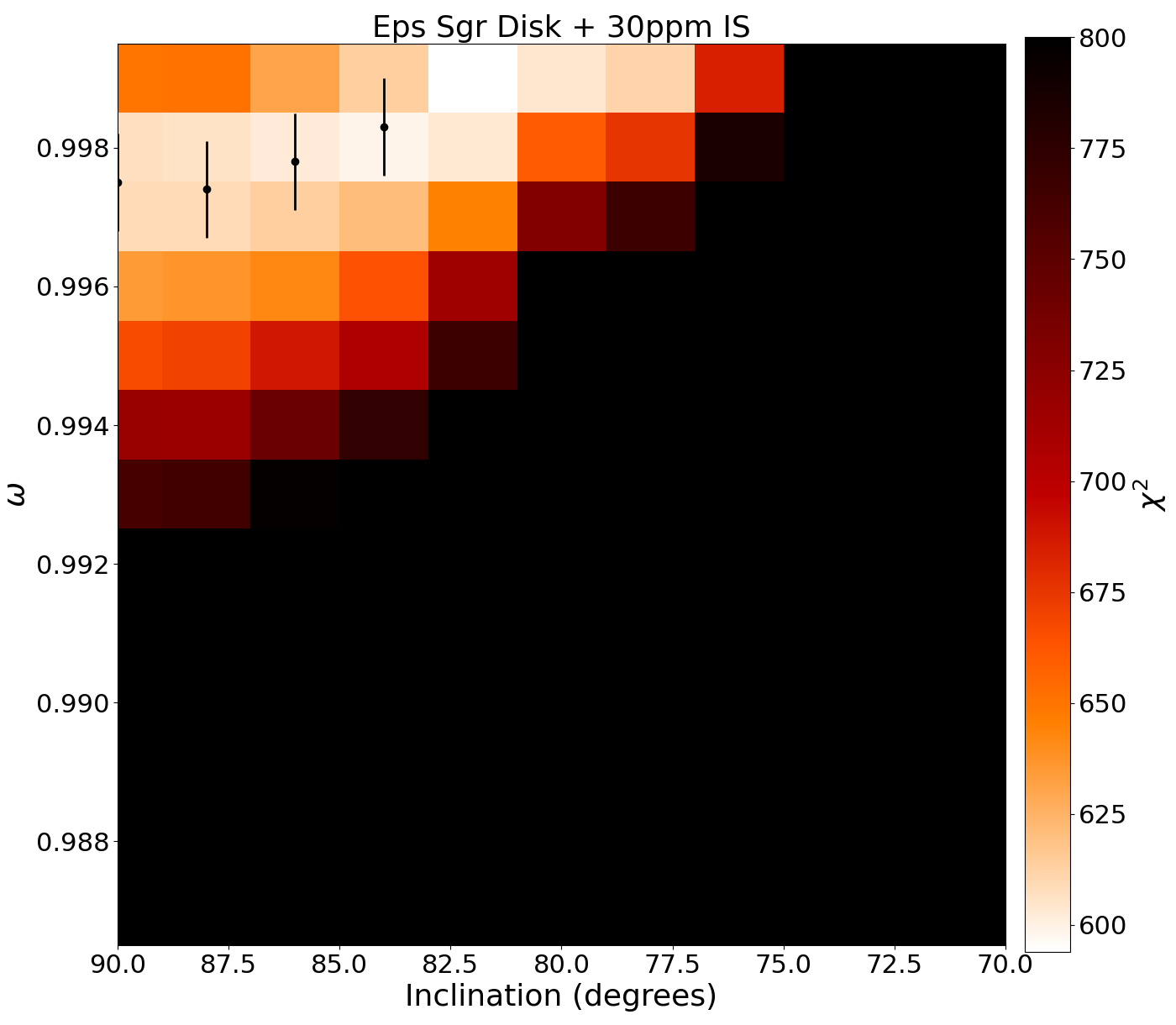}
    \caption{As Figure \ref{fig:fit_offset1} but using the rotational polarization models plus a disk and up to 30ppm interstellar component}
    \label{fig:fit_disk30}
\end{figure}

\begin{figure}
    \centering
    \includegraphics[width=\columnwidth]{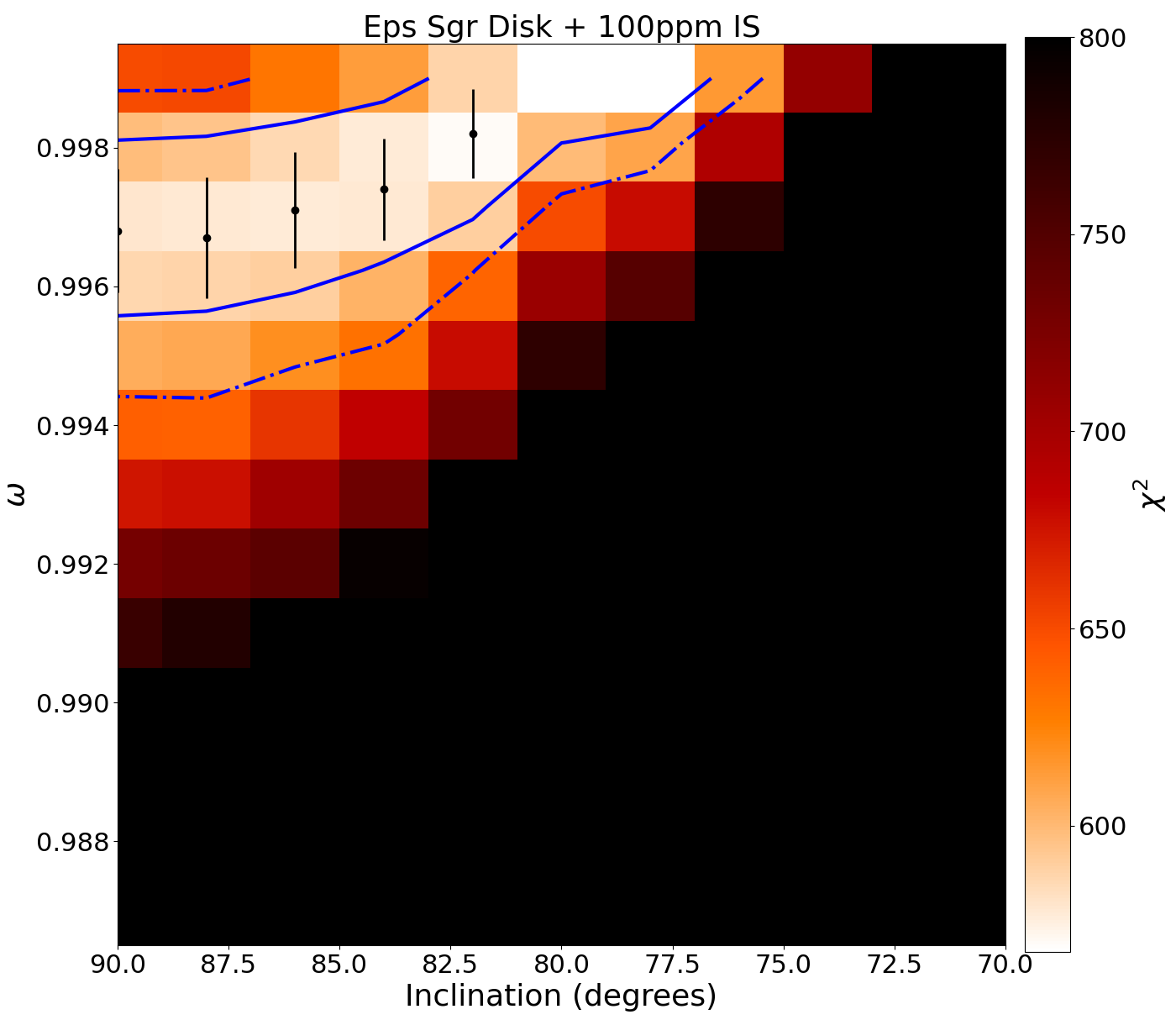}
    \caption{As Figure \ref{fig:fit_offset1} but using the rotational polarization models plus a disk and up to 100 ppm interstellar component. The solid lines show the 2$\sigma$ contours on the rotational parameters as fitted to these models (see Section \ref{sec:uncertainties}). The dash-dot lines shows the corresponding extended range of parameters when the uncertainties on the input parameters to the model grid (in particular $v_{\rm e} \sin{i}$) are taken into account.}
    \label{fig:fit_disk100}
\end{figure}

\begin{table}
    \caption{Parameters of Best Fit Region in Grid Model Variants (Figures \ref{fig:fit_offset1} to \ref{fig:fit_disk100})}
    \centering
    \tabcolsep 3.5 pt
    \begin{tabular}{lllll}  \hline
   Variant & QU Offset &  \multicolumn{3}{c}{Disk + Interstellar}  \\ \hline
   $p_{\rm max}$ (ppm) & 0.0 & 30.0 & 60.0 & 100.0   \\
   $\theta_{\rm ism}$ (deg)$^a$  &  &  144.3 & 137.1 &  134.2   \\
   $p_{\rm disk}$ (ppm)$^a$   &  276  &  299  & 330 & 367 \\
   $\theta_{\rm rot}$ (deg)   &  40.1  & 39.9 & 39.9 &  39.9 \\
   $\omega (i=90)$   &    0.9977(7) & 0.9975(7) & 0.9972(7) & 0.9968(9) \\
   $\omega (i=88)$   &    0.9977(7) & 0.9974(7) & 0.9971(7) & 0.9967(9) \\
   $\omega (i=86)$   &    0.9981(6) & 0.9978(7) & 0.9975(7) & 0.9971(8) \\
   $\omega (i=84)$   &    0.9986(6) & 0.9983(7) & 0.9979(6) & 0.9974(7) \\
   $\omega (i=82)$   &    0.999 & 0.999 & 0.9987(5) & 0.9982(6) \\
   $\omega (i=80)$   &    &  & 0.999 & 0.999 \\
\hline
    \end{tabular}
\begin{flushleft}
Notes: \\
Figures in parentheses are the errors in the last digit of the value. When the best fit for an inclination is in the $\omega$=0.999 bin the value give is the bin centre and no error can be determined. \\
$^a$  Varies slightly with inclination, average value is given. \\
\end{flushleft}
    \label{tab:fit_regions}
\end{table}

\begin{table}
   \caption{Best-fitting Stellar Parameters.}
    \centering
    \tabcolsep 8 pt
   \begin{tabular}{lrrr}
   \hline   Parameter  &  $i$=90$^a$ & Best$^a$ &   Range$^b$ (2$\sigma$)  \\ \hline
    $\omega$   &  0.997  &  0.999  &  0.995 -- 1.0  \\  
    $v_{\rm e}/v_{\rm crit}$  &  0.95 & 0.97  &  0.94 -- 1.0 \\
    Inclination, $i$ [\degr]     &    90 &  80  &  76 -- 90 \\
    $T_{\rm eff}$ [K]  &   10091  &  9950 &  9845 -- 10096 \\
    $T_p$ [K]  &   11791  &  11721 &  11548 -- 11813 \\
    $T_e$ [K]  &   7884  &  7433 &  7324 -- 7492 \\
    Log($L$/\lsun)   &  2.709 &  2.696  &  2.669 -- 2.726 \\
    R$_p$ [\rsun]  &  5.98 &  6.01 &  5.90 -- 6.08  \\
    R$_e$ [\rsun]  &  8.59 &  8.80  &   8.47 -- 8.89 \\
    Mass [\msun]  &  3.69 & 3.80 &  3.69 -- 3.90  \\
    $\log{g_p}$ [dex cgs]  &  3.45 &  3.46 &  3.45 -- 3.48  \\
    $\log{g_e}$ [dex cgs]  &  2.24 &  2.00 &  1.99 -- 2.36  \\
    $P_{\rm rot}$ [days] & 1.63 & 1.61 & 1.57 -- 1.64 \\
    
    \hline
    \end{tabular}
\begin{flushleft}
Notes: \\
$^a$  The parameters listed are those of the grid model (see Section \ref{sec:grid}) giving the lowest $\chi^2$ among all models (Best), and the best model at inclination of 90 degrees. \\
$^b$ Range is the full range of stellar parameters corresponding to the region enclosed by the dot-dash line in Figure \ref{fig:fit_disk100}.
\end{flushleft}
    \label{tab:fit_pars}
\end{table}

\subsection{Comparison with Observations}

In comparing the model grid calculations with the observations we have to contend with the variability of the system (Section \ref{sec:variability}) and the presence of additional polarization components due to the disk and interstellar polarization (Section \ref{sec:components}). There is insufficient data to attempt to correct for the variability. However, examination of the polarization measurements shows that the earliest observations (those in 2014 to 2016) seem to show the largest scatter, and variability at these times is confirmed by the spectra shown in Figure \ref{fig:uves_cfht}. We therefore chose to exclude these observations from the analysis, which was restricted to the 22 observations covering 2017 -- 2023. There is still some evidence of variability, which shows up as excess noise compared with that expected from the measurement errors. 

The additional polarization components were handled by fitting the observations with several variants of the model. The same grid of calculations was used for the polarization of the distorted star, but different models were used for the polarization of the additional components. Initial tests over a wider and coarser grid of models showed that only the models with high $\omega$ were compatible with the polarization. The final grid covered the range 0.980 $\leq \omega \leq$ 0.999 and 70 $\leq i \leq$ 90 and the relevant section of this region is shown in Figures \ref{fig:fit_offset1} to \ref{fig:fit_disk100}.

In the first variant (Figure \ref{fig:fit_offset1}) we represent the additional polarization as a fixed (i.e. wavelength independent) offset in Q and U. Thus at each point in the grid we fit three values, the Q-offset, U-offset and the position angle of the star's rotation axis ($\theta_{\rm rot}$). With this model we find that for the best-fitting (lowest-$\chi^2$) points in Figure~\ref{fig:fit_offset1}, the additional QU-offset corresponds to a position angle of 38.5$\degr$, and the star's rotation axis is at 40.1$\degr$. The polarization is within 1.6$\degr$ of that expected for an equatorial disk (Section \ref{sec:components}). The small difference in angles can be understood as due to an additional contribution from interstellar polarization.

In the remaining variants we represent the additional polarization as a combination of a 
wavelength-independent disk contribution, and an interstellar-polarization component following the `Serkowski law' \citep{Serkowski1973,serkowski75,wilking82,whittet92}. We set the $\lambda_{\rm max}$ of the interstellar polarization to 470 nm, the typical value found within the Local Hot Bubble (\citealp{marshall20} and refs. within). In these models there are four fitted parameters at each grid point:  $p_{\rm max}$ and the position angle $\theta_{\rm ism}$ of the interstellar polarization, the position angle $\theta_{\rm rot}$ of the star's rotation axis, and the polarization of the disk $p_{\rm disk}$. The disk is assumed to be in the equatorial plane so its polarization position angle is $\theta_{\rm rot}$. 

There is some degeneracy in these models between $p_{\rm max}$ and $p_{\rm disk}$. We therefore ran three versions of the models where we applied the restricted ranges $0 < p_{\rm max} \leq 30$~ppm, 
$0 < p_{\rm max} \leq 60$ ~ppm, and $0 < p_{\rm max} \leq 100$~ppm, based on our analysis of the likely value of interstellar polarization as described in Appendix \ref{apx:interstellar}. Although we constrain $p_{\rm max}$ to a range, the best-fitting models in all these cases have $p_{\rm max}$ at the largest allowed value. The results for two of these cases are plotted in Figures $\ref{fig:fit_disk30}$ and $\ref{fig:fit_disk100}$, and all the variants are described in \mbox{Table \ref{tab:fit_regions}}. 

All these variants (Figures ~\ref{fig:fit_offset1}--\ref{fig:fit_disk100}) produce quite similar results, with a ``valley'' of low $\chi^2$ values running from $\omega \sim 0.997$ at $i = 90\degr$ to $\omega = 0.999$ at $i \sim 80\degr$. Thus despite different assumptions about the interstellar polarization, almost the same stellar parameters are being selected in all these cases.

In Table \ref{tab:fit_pars} we list the stellar parameters corresponding to the two grid models at the  ends of this $\chi^2$ valley. The model labelled ``Best'' (in the `Disk + 100 ppm' set) is the one that has the lowest $\chi^2$ and is at $\omega$ = 0.999, $i$ = 80$\degr$ (in some of the other model variants the best fit is in the adjacent grid point at $i$ = 82$\degr$). However, the very strong shell absorption seen in $\epsilon$ Sgr, despite the low density of the disk, might favour a higher-inclination model like the $\omega$ = 0.997, $i$ = 90$\degr$ case also listed in Table $\ref{tab:fit_pars}$.

Figure \ref{fig:model_image} shows images of the stellar models for the two cases listed in Table \ref{tab:fit_pars}. The models show the distribution of specific intensity over the star, showing the strong effect of gravity darkening at the equator, and is overlaid with polarization vectors. The polarization is dominated by light from the bright regions at top and bottom of the image resulting in the negative (perpendicular to the rotation axis) integrated polarization at this wavelength.

\begin{figure}
    \centering
    \includegraphics[width=\columnwidth]{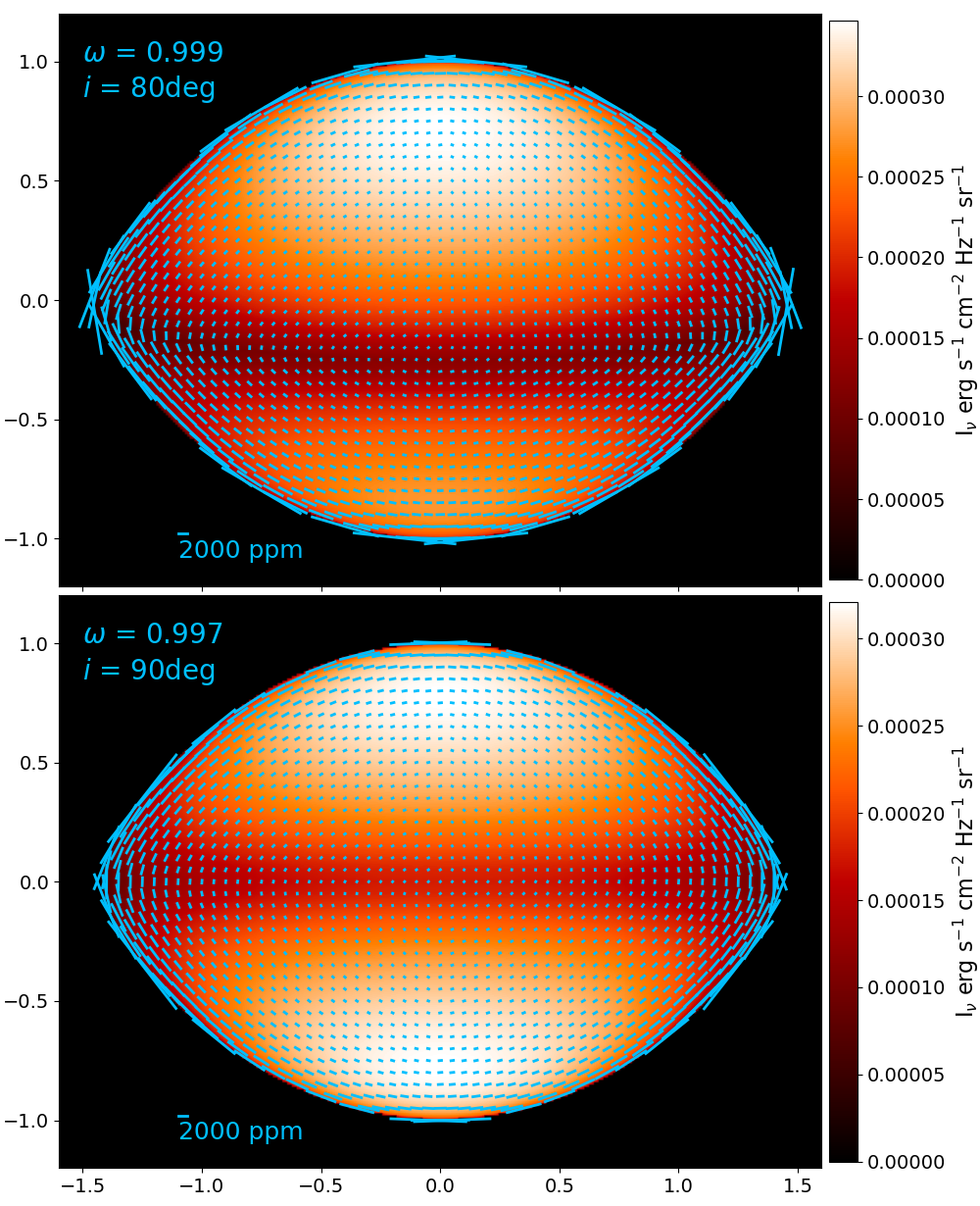}
    \caption{Polarization and intensity images for the two models of $\epsilon$ Sgr listed in Table \ref{tab:fit_pars}. Specific intensity is overlaid with polarization vectors for a wavelength of 400 nm.}
    \label{fig:model_image}
\end{figure}

\subsection{Uncertainties}
\label{sec:uncertainties}

To determine the uncertainties on the fitted values of $\omega$ and $i$ we use the bootstrap method \citep{Numerical_recipes}. For each of the model grid variants we ran the modelling on 1000 trials using random selections of the 22 observations with replacement (so that each observations can be selected more than once). For each trial we determined the best $\omega$ value at each inclination by fitting the $\chi^2$ values with a cubic function. We then determined the standard deviation of these values over the 1000 trials. These results give the errors on $\omega$ listed in Table \ref{tab:fit_regions} and plotted as error bars on Figure \ref{fig:fit_offset1} to \ref{fig:fit_disk100}. This method of error determination does not depend on the errors on the individual polarization measurements, but on the actual scatter on the data, which is largely due to the polarization variability of the star.

Using these bootstrap results we can determine the $\Delta\chi^2$ ranges corresponding to 1 or 2$\sigma$, and thus determine the 2-$\sigma$ contours on $\omega$ and $i$ which are shown as the solid lines in Figure~\ref{fig:fit_disk100}. We use Figure~\ref{fig:fit_disk100} (the `Disk + 100 ppm' variant) as this gives the lowest $\chi^2$ values, and has a somewhat broader distribution in $\omega$, but largely overlays the best-fit regions covered by the other model variants.

Additional uncertainties are introduced by the uncertainties in the input parameters to the model grid (Section \ref{sec:grid}). These were assessed by running the modelling for auxiliary grids calculated for adjusted values of $v_{\rm e}\sin{i}$, parallax and UV flux in turn. The most significant effect is that due to an estimated 10 km s$^{-1}$ uncertainty in $v_{\rm e}\sin{i}$ which has the effect of moving the $\chi^2$ valley up and down in $\omega$. This leads to the wider 2-$\sigma$ contours shown by the dash-dot lines in Figure \ref{fig:fit_disk100}. Uncertainties in parallax and UV flux do not significantly change the fitted values of $\omega$ and $i$, but do have a small effect on the relationship to other stellar parameters. The wider contours (dash-dot line) were then used to determine the allowed range of each stellar parameter, as listed in Table \ref{tab:fit_pars}.

\needspace{2\baselineskip}
\section{Discussion}

\subsection{There is No Polarizing Debris Disk}

Previously, HIPPI observations of $\epsilon$ Sgr were reported by \citet{cotton16c}. They interpreted the observed polarization in the context of the system's reported infrared excess \citep{rodriguez12}. Cotton et al. present a model comprising two dust components (to reproduce the continuum IR excess) in addition to the stars. A cold dust belt (responsible for the bulk of the IR excess) was assumed to be circumbinary, with a warm dust belt orbiting one of the two stars. A residual polarization of 30~ppm remained after accounting for polarization induced by the interstellar medium. To reproduce this a fractional polarization of around 50\% of the total IR excess would be required. Such a value is much higher than those measured for dust grains in other debris disks, which peak at around 20 to 30\% (e.g. \citealp{krivova00, graham07, anche23}). Since the \citet{cotton16c} model was proposed, work by \citet{vandeportal19} has shown that warm (mid-IR) excesses represent substantially lower dust masses and contribute less to polarization compared to cool (far-IR) excesses with lower fractional luminosities but representing greater dust masses. This further reduces the plausibility of the proposed disk model.

The large beam size of IRAS and the low galactic latitude of the system ($|b| < 10\degr$) also lead to the possibility that the flux density measurement at 60~$\mu$m is contaminated by diffuse background emission. Indeed, the IRAS point source catalog \citep{iras_psc94} notes a quality flag of ``2'' for the 60~$\mu$m band. Applying a downward revision of the flux, as implied, would only increase the tension between the level of IR excess and the polarization efficiency of the disk dust required.

In the light of observations presented here we can eliminate the disk model as a plausible explanation for the polarimetric signal arising from the system. Specifically, the invariance of the position angle of the polarimetric signal between epochs, despite the secondary's proper motion, is inconsistent with illumination of a circumbinary dust disk. The agreement between the $r^\prime$ observations with small and large aperture sizes in the 2018MAR run is also inconsistent with this hypothesis. Furthermore, the level of fractional polarization required would be a factor of two higher than other systems for which such measurements are available (see above, and also \citealp{marshall23}). 

Given that \citet{rodriguez12} find the system to be unstable with the accepted debris disk architecture and that IR excess may also be produced by free-free emission in the (previously unidentified) Be disk \citep{kastner89}, it is likely no debris disk exists at all. The nature of $\epsilon$ Sgr as a rapidly rotating B star with a gas disk is therefore the origin of its polarization. 

\subsection{Rapid Rotation}

The rotation as a fraction of criticality that we measure for $\epsilon$ Sgr of $\omega$ $\sim$ 0.997, which corresponds to $v_{\rm e}/v_{\rm crit}$ $\sim$0.95, is the largest yet measured for a star in our galaxy\footnote{VFTS285, an O star in the Large Magellanic Cloud, has been reported as having $\omega$ = 1.00 \citep{shepard22}}. The previous highest was obtained by \citet{desouza14} who measured the rotation of Achernar using interferometry as $\omega$ = 0.980, $v_{\rm e}/v_{\rm crit}$ = 0.883. Achernar is a Be star, but the measurements were made during a quiescent phase where the Be disk was absent. 

$\epsilon$ Sgr stood out in our polarization study of bright, rapidly-rotating stars because of the large amplitude of the wavelength-dependent polarization variations -- $>$300 ppm, compared with $<$100 ppm for the other rapidly-rotating stars \citep{cotton17,bailey20b,howarth23}. The only other star with a comparable polarization amplitude was $\theta$ Sco \citep{lewis22}. In that case the rotation was slower than in $\epsilon$ Sgr, but the low gravity of this evolved star enhanced the polarization.

Multiwavelength polarimetry should therefore be a good way of finding other similar extreme rotators, and establishing how common such objects are. In the 19 stars we have observed for our AAT polarimetry program on rotating stars we have found no other candidates for such extreme rotation. The program is based on stars down to V $\sim$ 3.5 from the list of \cite{vanbelle12}.

\begin{figure*}
    \centering
    \includegraphics[width=18cm]{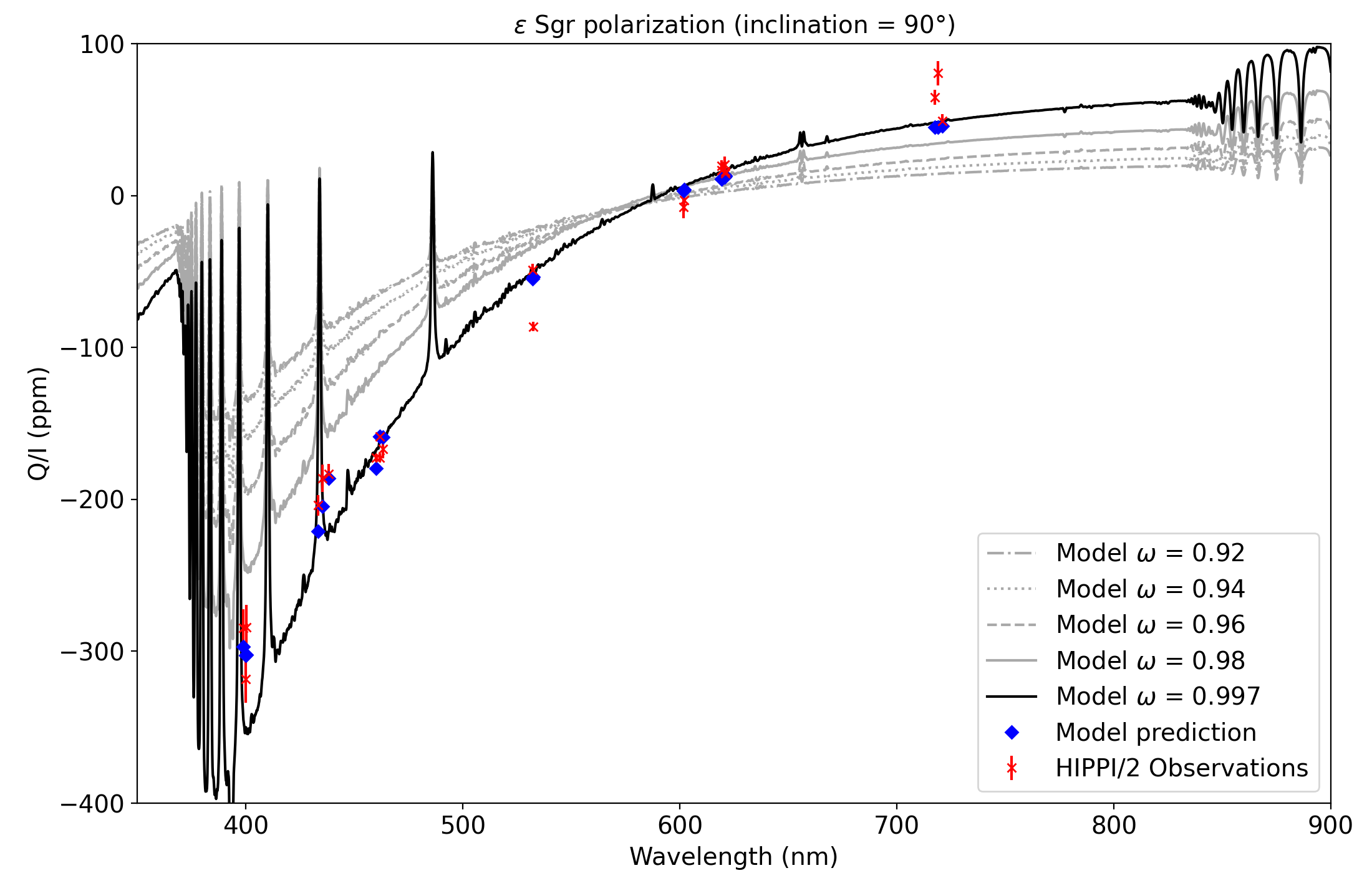}
    \caption{Comparison of observations and models for the polarization of $\epsilon$ Sgr as a function of wavelength. The models show the predictions due to stellar rotation. The disk and interstellar polarization have been applied as a correction to the observations. The black solid line is the model for $\omega$ = 0.997, $i$ = 90$\degr$. Blue diamonds are the predicted polarization values for each observed filter (they are not necessarily exactly on the line, because they are averages over the filter bandpass). Grey lines show the model predictions for grid models with lower values of $\omega$.}
    \label{fig:omega_dep}
\end{figure*}

\begin{figure}
    \centering
    \includegraphics[width=\columnwidth]{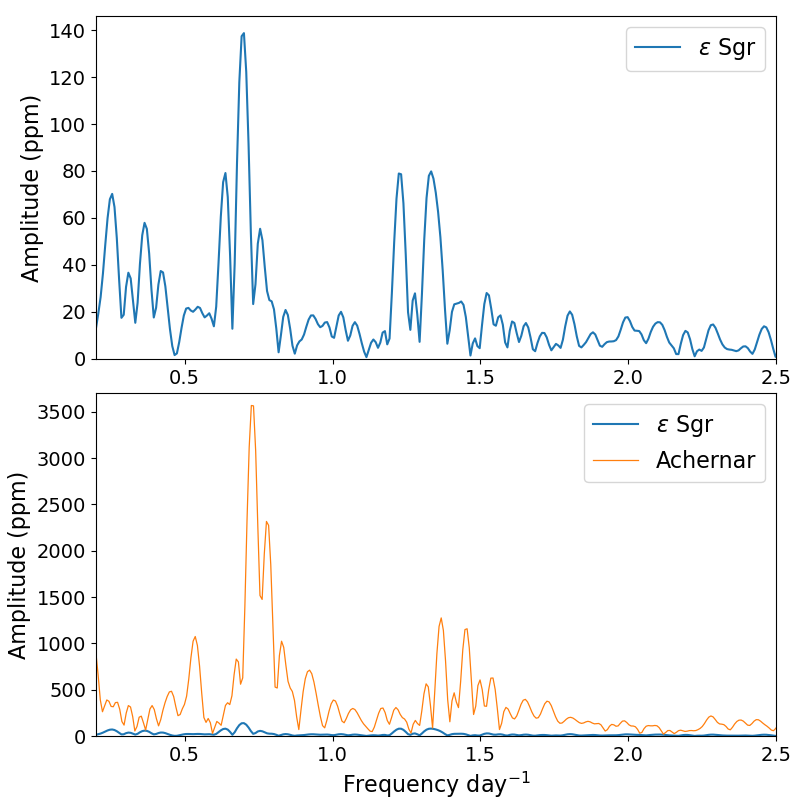}
    \caption{Periodogram of TESS light curve of $\epsilon$ Sgr (TESS sector 13). In the upper panel it is shown at full scale. In the lower panel it is compared with the periodogram of the bright classical Be star Achernar ($\alpha$ Eri) from TESS sector 2. The presence of two frequency groups with a 2-to-1 frequency ratio is common in Be stars, but the amplitude seen in $\epsilon$ Sgr is much lower than is commonly seen.}
    \label{fig:tess}
\end{figure}

\subsection{Relationship to Be Stars}

The value of $v_{\rm e}/v_{\rm crit}$ $\sim$0.95 that we find for $\epsilon$ Sgr is interesting because this is the value that was suggested \citep{townsend04,owocki05} as required for ``weak'' processes such as pulsation to be able to launch material from the star into orbit. If all Be stars rotated this fast, the existence of Be disks would be easier to explain. However, statistical analysis of $v_{\rm e} \sin{i}$ data for Be stars (\citealt{cramner05} [but see \citealt{Howarth07}]; \citealt{zorec16}) suggests that most Be stars rotate much slower than this, with a broad distribution of $v_{\rm e}/v_{\rm crit}$ values from $\sim$0.5 to $\sim$0.95, with the mode of the distribution at 0.66 \citep{zorec16}. There are also rapidly-rotating B stars \citep[e.g. Regulus,][]{cotton17} that are not Be stars. 

While $\epsilon$ Sgr has a rotation rate at the top end of the range inferred for Be stars it is certainly not a typical classical Be star. It does appear to have an equatorial disk, but it is very weak compared with typical Be stars. Weak H$\alpha$ emission has been observed on one occasion but is normally absent, as discussed in section \ref{sec:variability}. Thus, these results confirm that rotation alone cannot be the key requirement for formation of Be disks. Rather, there must be some additional mechanism operating in Be stars that can launch material from the surface into orbit. In $\epsilon$ Sgr this mechanism is either not operating or is operating with much reduced efficiency, enabling the formation of only a weak disk despite the very rapid rotation. We note that Classical Be stars are more common at early B spectral types and less common at the B9/A0 spectral type of $\epsilon$ Sgr. \citet{Huang10} suggest that the threshold rotation rate for forming a Be disk is much higher for later type stars.

Most classical Be stars show photometric variability on short timescales. Observations with the Transiting Exoplanet Survey Satellite \citep[TESS,][]{ricker15} typically show multiple periodic signals at frequencies between 0.5 and 4 day$^{-1}$ \citep{labadie-bartz22}. These are normally interpreted as non-radial pulsations, which may be related to the mass ejection process that forms the disk \citep{rivinius13}.

In Figure \ref{fig:tess} we show the periodogram of the TESS two-minute cadence light curve of $\epsilon$ Sgr for sector 13 (2019 Jun 19 --- Jul 17)\footnote{The TESS data used here are available at MAST: \dataset[10.17909/t1ae-rd45]{\doi{10.17909/t1ae-rd45}}}. The pattern seen of two groups of peaks (here at $\sim$0.65 and 1.3 days$^{-1}$), with a frequency ratio of roughly 1 to 2 is one that is very commonly seen in Be stars \citep[e.g.][]{walker05,semann18,baade18,labadie-bartz22}. However the amplitude (around 140 ppm for the highest peak) is low compared with those more typically seen in classical Be stars, which are in the parts per thousands \citep{labadie-bartz22}.

As an example we show in the lower panel for Figure \ref{fig:tess} the periodogram of $\epsilon$ Sgr compared with that of the bright Be star Achernar ($\alpha$ Eri, HD 10144), based on observations obtained from TESS sector 2 (2018 Aug 23 --- Sep 20). At that time Achernar was in an active state. It has been quiescent without a disk since late 2019\footnote{According to our polarization monitoring of Achernar over the last 10 years.}. The frequency structure seen in Achernar is quite similar to that of $\epsilon$ Sgr, but the peak amplitude is 25 times higher in Achernar. 

\bigbreak
\bigbreak
\subsection{Evolutionary Status}

\begin{figure}
    \centering
    \includegraphics[width=\columnwidth]{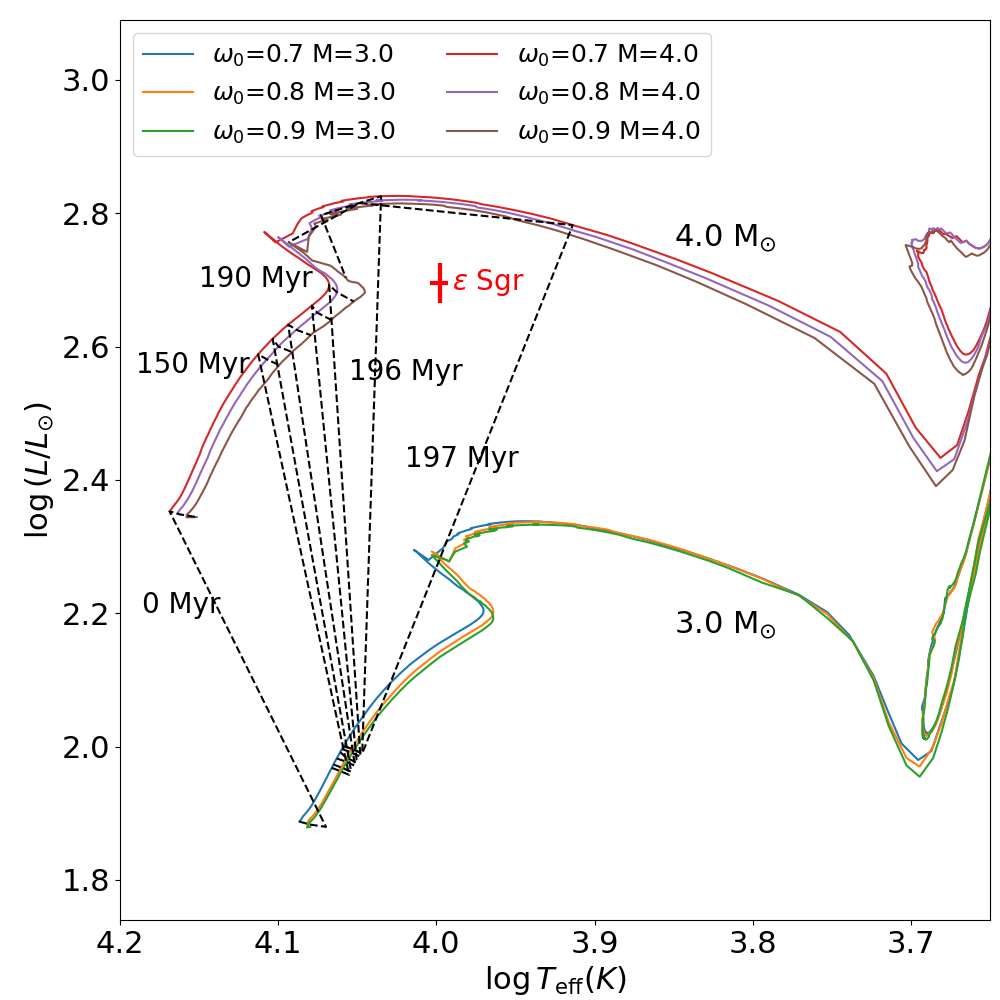}
    \caption{Location of $\epsilon$ Sgr on the HR diagram compared with evolutionary models for rotating stars from \citet{georgy13}. Isochrones are shown at 10 Myr intervals from 150 to 190 Myr, and then at 196 and 197 Myr. }
    \label{fig:evol_hr}
\end{figure}

\begin{figure}
    \centering
    \includegraphics[width=\columnwidth]{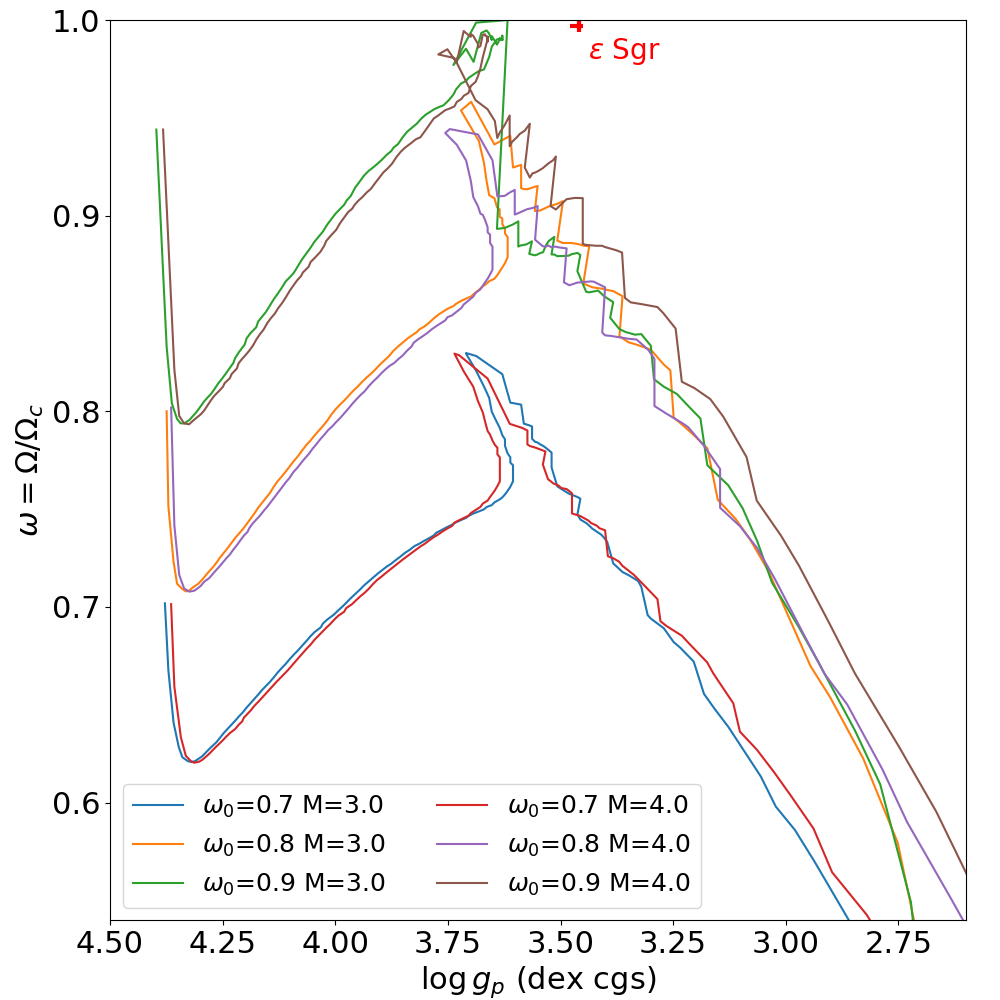}
    \caption{Plot of $\omega$ against $\loggp$ for $\epsilon$ Sgr and the same models as in Figure \ref{fig:evol_hr}. The rotation rate ($\omega$) reaches a maximum at a point corresponding to the ``hook'' in the HR diagram, and then declines rapidly as the star evolves across the Hertzprung gap.}
    \label{fig:evol_gw}
\end{figure}

In Figure \ref{fig:evol_hr} we show the position of $\epsilon$ Sgr on an HR diagram compared with evolutionary models for rotating stars of solar metallicity (Z = 0.014) from \cite{georgy13}. The models are for $M/\msun$ of 3.0 and 4.0 (our measured value is $M/\msun$ = 3.69), and initial rotation rates of $\omega$ = 0.7, 0.8 and 0.9. The location of $\epsilon$ Sgr is that of a star just beginning to evolve away from the main sequence. Its position corresponds to the hydrogen shell burning phase where the star is crossing the Hertzprung gap. This is a very rapid phase of stellar evolution and such stars are not commonly observed. 

Figure \ref{fig:evol_gw} shows the comparison with the same set of models in a plot of $\omega$ against \loggp. The \loggp\ we measure for $\epsilon$ Sgr is again consistent with a star beginning to evolve away from the main sequence. The maximum rotation rate for these model stars corresponds to the ``hook'' in the HR diagram where the star contracts after the end of core hydrogen burning. At this point a star born with $\omega$ = 0.9 or greater reaches near critical rotation. However, Figures \ref{fig:evol_hr} and \ref{fig:evol_gw} show that our measured parameters for $\epsilon$ Sgr place the star at a point beyond the ``hook'' where the star is expanding and the rotation rate is slowing. The rotation we measure for $\epsilon$ Sgr is therefore too high to match any of these single rotating star models. Its rapid rotation may therefore be the result of close binary evolution \citep{pols91}.

The visible companion (Section \ref{sec:imaging}) seems to have colours consistent with a main sequence star making it unlikely to have been involved in the interaction that caused the rapid rotation. $\epsilon$ Sgr is known to be an \mbox{X-ray} source \citep{berghoefer96,hubrig01}, which might indicate an unseen companion. 

\section{Future Observations}

Our results on $\epsilon$ Sgr could be further tested by a number of other observations.

\subsection{Interferometry}

Interferometry has been successful in imaging the rotational distortion of a number of stars of similar brightness and spectral type to $\epsilon$ Sgr such as Regulus \citep{che11}, Altair \citep{monnier07} and $\alpha$ Oph \citep{zhao09}. Such observations should be able to confirm the extreme rotational oblateness and detect the variation of surface brightness due to gravity darkening. Interferometry is better than polarimetry at measuring the inclination and hence distinguishing between the range of models that are consistent with the polarimetry.

\begin{figure}
    \centering
    \includegraphics[width=\columnwidth]{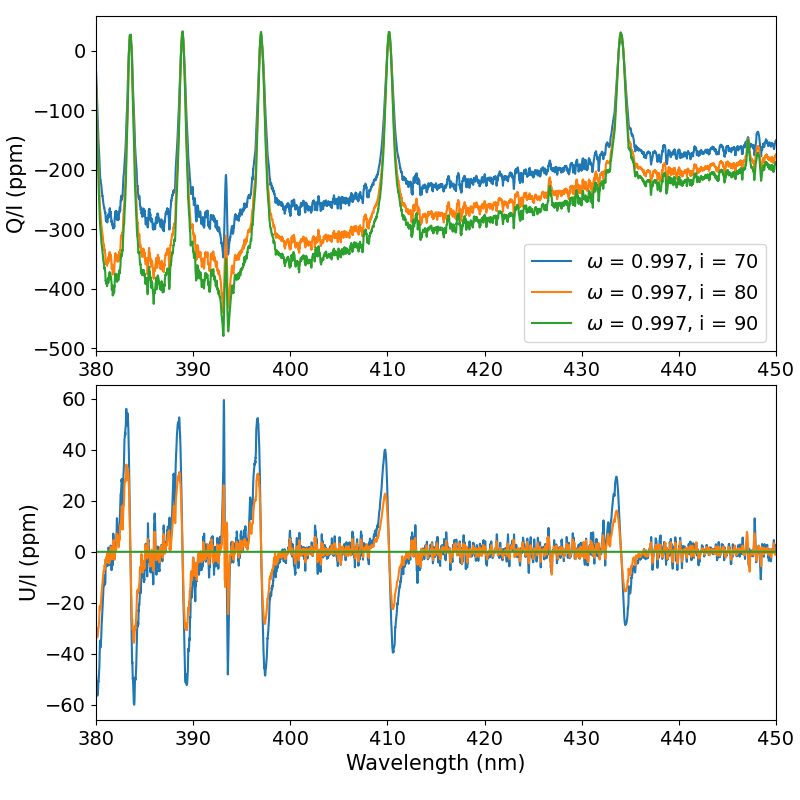}
    \caption{Modelled polarization for $\epsilon$ Sgr at high spectral resolution. The pure \"{O}hman effect (due to rotational line broadening) is seen most clearly on the CaII K line at 393 nm, the strongest metal line in this region. On the strongly Stark broadened hydrogen lines (the five strongest lines in this region) the \"{O}hman effect in Q/I is swamped by other effects, but the \"{O}hman effect is still present in U/I, and is a strong function of inclination}
    \label{fig:ohman}
\end{figure}

\subsection{Spectropolarimetry and the \"{O}hman Effect}

Current observations of the polarization due to the rotational distortion of stars, such as those presented in this paper, have been made with broad-band polarimeters because these instruments can achieve the high precision needed to measure the small rotational signals. However, further information on stellar rotation could be obtained from spectropolarimetric observations with sufficient resolution to resolve the lines.

These effects include the \"{O}hman effect first predicted in 1946 \citep{ohman46}. The \"{O}hman effect arises because at different points in the line profile of a rotationally broadened line the absorption arises in different parts of the star and therefore changes the integrated polarization we observe. Models of the effect have been given by \citet{collins91b}. The effect is small and has not yet been observed. Since $\epsilon$ Sgr has the largest rotational signal yet seen in continuum polarization it is interesting to consider whether observations of the \"{O}hman effect for this star are feasible and useful.

Because our polarization modelling using SYNSPEC/VLIDORT (Section \ref{sec:polmod}) is carried out at high spectral resolution, our models predict what should be seen in spectropolarimetry. As in the modelling done by \citet{collins91b} the spectral lines are treated as pure absorbers. No line polarization processes are involved. The polarization is due to the same processes that produce the continuum polarization (primarily electron scattering).

Models for the polarization wavelength dependence of $\epsilon$ Sgr are shown in Figure \ref{fig:ohman} for the 380 -- 450 nm region. This shows the problem with detecting the \"{O}hman effect in a star of this spectral type. The spectrum is dominated by hydrogen lines, which are strongly Stark broadened by an amount that exceeds the rotational broadening. The pure \"{O}hman effect is only seen on the metal lines, of which the CaII K line at 393 nm is the strongest in this region. The large polarization features seen on the hydrogen lines in Q/I are not due to the \"{O}hman effect, but are due to the fact that in these strong lines we are seeing different layers of the star in line and continuum.

The \"{O}hman effect features seen in U/I are more interesting because they show a strong dependence on inclination. The features seen on the hydrogen lines are smeared out by the Stark broadening, but may still be observable and would provide another way of refining the inclination of the star.  Such an observation would require spectropolarimetry with R $>$ 5000 and precision $\sim$ 5 ppm or better.

\subsection{Ultraviolet (UV) Polarimetry}

Models of rotating stars predict that polarization values are expected to increase at shorter wavelengths \citep{collins91a,jones22} reaching $\sim$1\% (10000 ppm) at 120 nm for large $\omega$ and high inclination. Since the polarization due to a disk is not strongly wavelength dependent, and interstellar polarization decreases at shorter wavelengths, observations in the UV would remove any possible confusion between these different sources. Such observations may be possible with future space missions similar to the proposed Polstar satellite \citep{scowen22}.

\section{Conclusions}

Multi-wavelength high-precision polarization observations of $\epsilon$ Sgr show the distinctive wavelength dependence expected for a rotationally distorted star, but with a much higher amplitude ($\sim$ 300 ppm) than seen in other rapidly rotating stars. In order to fit a rotating-star model, an additional positive polarization component is needed. We attribute this to polarization in a low-density ($\sim$10$^{-13}$ g cm$^{-3}$) equatorial gas disk which also produces the narrow shell absorption features seen in the spectrum. 

Detailed modelling of the polarization together with additional data including the spectral-energy distribution and $v_{\rm e} \sin{i}$ reveals a very high rotation rate of $\omega$ $\geq$ 0.995, the highest yet measured for a star in our galaxy. The modelling also results in a detailed set of stellar parameters. The star is rotating too rapidly to be consistent with single rotating star evolutionary tracks, suggesting that it is likely to be a product of binary evolution.

$\epsilon$ Sgr shares some features with Be stars in having a low density gas disk, weak H$\alpha$ emission seen on one occasion, and multiple periodicities in its TESS light curve. However, the disk is of very low density and the TESS periodicites are of very low amplitude compared with typical Be stars. The results confirm that rapid rotation alone cannot be the cause of the Be phenomenon, and support the idea that an additional mechanism is needed to eject material into a Be disk.


$\epsilon$ Sgr is an excellent target for future observations by interferometry, spectropolarimetry and UV polarimetry which would allow the extreme rotation to be confirmed and the stellar parameters refined.

\section{Acknowledgments}.
\begin{acknowledgments}
Based in part on data obtained at Siding Spring Observatory. We acknowledge the traditional owners of the land on which the AAT stands, the Gamilaraay people, and pay our respects to elders past and present. JPM acknowledges research support by the National Science and Technology Council of Taiwan under grant NSTC 112-2112-M-001-032-MY3. DVC thanks the Friends of MIRA for their support. 
\end{acknowledgments}

%

\vspace{5mm}
\facilities{AAT (HIPPI, HIPPI-2, IRIS2), CFHT (ESPaDOnS), VLT:Kueyen (UVES), IUE, TESS}





\appendix

\section{Interstellar Polarization}
\label{apx:interstellar}

\begin{figure*}
    \includegraphics[width=\textwidth]{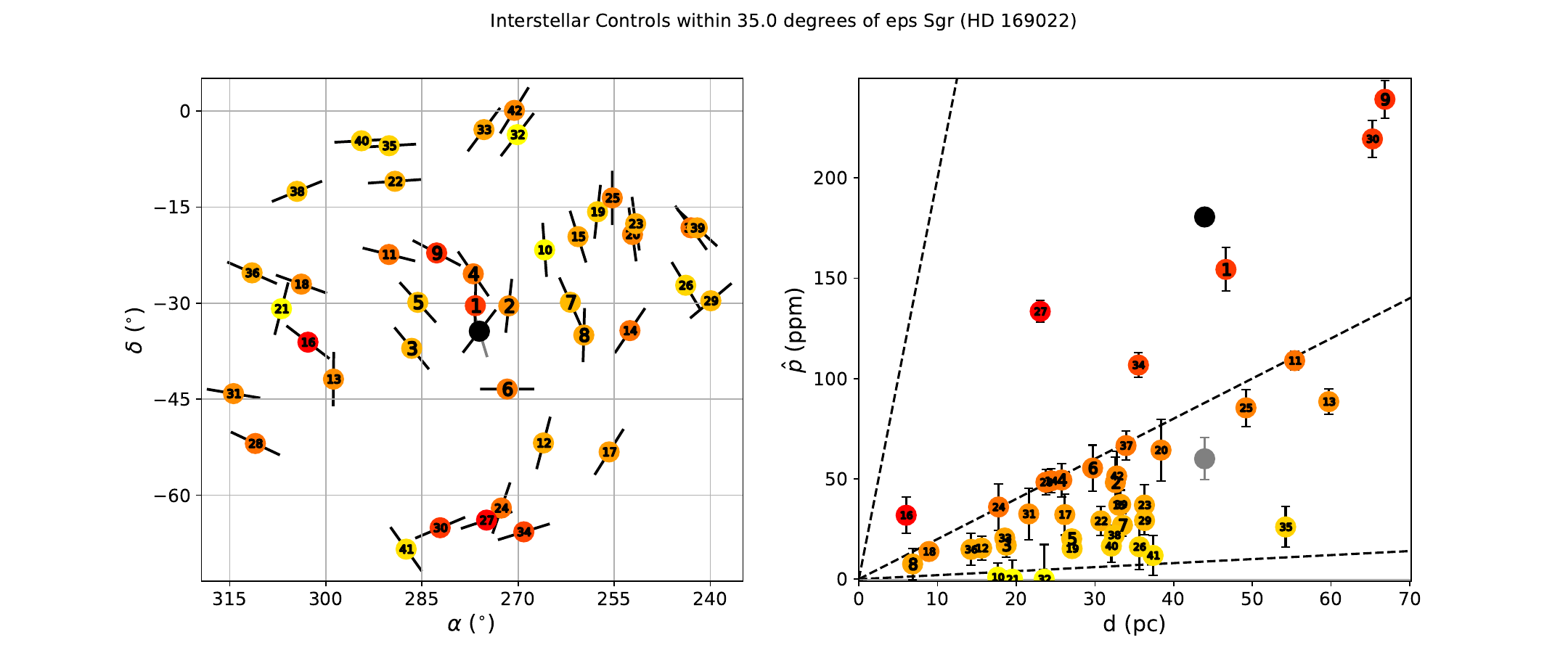}
    \caption{Plot of the polarization of stars near to $\epsilon$~Sgr, representative of interstellar polarization, as projected onto the sky (left) -- vectors represent the PA (clockwise N over E) -- and debiased according to $\hat{p}=\sqrt{p^2-p_e^2}$ as a function of distance (right). Where multiple observations exist, the error-weighted average is displayed. The grey vector and dot represent the interstellar polarization of $\epsilon$~Sgr as predicted by the model in \citet{cotton17b}, whereas their black counterparts average the observations, and correct them for wavelength as if they are interstellar measurements. All observations are scaled to 450~nm as if they are purely interstellar with $\lambda_{\rm max}$=470~nm. The controls are numbered in order of angular separation as follows, 1: HD\,169586, 2: HD\,165135, 3: HD\,177474, 4: HD\,166916, 6: HD\,176687, 7: HD\,165189, 8: HD\,157919, 8 HD 156384, 9: HD\,174309, 10: HD\,160915, 11: HD\,181240, 12: HD\,160691, 13: HD\,188114, 14: HD\,151680, 15: HD\,157172, 16: HD\,191408, 17: HD\,153580, 18: HD\,192310, 19: HD\,155125, 20: HD\,151504, 21: HD\,194640, 22: HD\,180409, 23: HD\,151192, 24: HD\,165499, 25: HD\,153631, 26: HD\,146070, 27: HD\,167425, 28: HD\,197157, 29: HD\,143114, 30: HD\,173168, 31: HD\,199288, 32: HD\,164259, 33: HD\,168723, 34: HD\,162521, 35: HD\,181391, 36: HD\,197692, 37: HD\,145518, 38: HD\,192947, 39: HD\,144766, 40: HD\,185124, 41: HD\,177389, 42: HD\,164651.  Note that for stars with $p < \sigma_{p}$ the PA's in the left hand panels are not well defined, since $\sigma_{\rm PA} > 28\degr$.}
    \label{fig:interstellar}
\vspace{-6 pt}
\end{figure*}

\begin{table*}
\caption{Observations of Interstellar Control Stars}
\begin{flushleft}
\label{tab:controls}
\tabcolsep 1.45 pt
\begin{tabular}{rlllrrcrrrrrr}
\hline
\multicolumn{1}{c}{Control}      & SpT   &   \multicolumn{1}{c}{Run}     & \multicolumn{1}{c}{UT}                  & Dwell & Exp. & \multicolumn{1}{c}{$\lambda_{\rm eff}$} & Eff. & \multicolumn{1}{c}{$q$} & \multicolumn{1}{c}{$u$} & \multicolumn{1}{c}{$p$} & \multicolumn{1}{c}{$\theta$}\\
\multicolumn{1}{c}{HD} &        &            &           & \multicolumn{1}{c}{(s)} & \multicolumn{1}{c}{(s)} & \multicolumn{1}{c}{(nm)} & \multicolumn{1}{c}{(\%)} & \multicolumn{1}{c}{(ppm)} & \multicolumn{1}{c}{(ppm)} & \multicolumn{1}{c}{(ppm)} & \multicolumn{1}{c}{($^\circ$)}\\
\hline
157919 & F5III-IV & 2018MAR & 2018-03-27 19:22:38 & 1114 &  640 &  468.3 &   84.2 &    18.5 $\pm$    \phantom{0}5.8 &$-$20.4 $\pm$    \phantom{0}5.6 &   27.5 $\pm$    \phantom{0}5.7 &  156.1 $\pm$   \phantom{0}6.0 \\
160691 & G3IV-V   & 2018MAR & 2018-04-04 17:15:17 & 1780 & 1280 &  472.1 &   85.7 &    14.5 $\pm$    \phantom{0}6.0 &    8.1 $\pm$    \phantom{0}5.9 &   16.6 $\pm$    \phantom{0}5.9 &   14.6 $\pm$   11.2 \\
165189 & A6V      & 2018AUG & 2018-09-01 10:51:27 &  971 &  640 &  464.4 &   58.7 & $-$56.6 $\pm$   11.7 &    0.1 $\pm$   11.2 &   56.6 $\pm$   11.5 &   89.9 $\pm$    \phantom{0}5.9 \\
168723 & K0III-IV & 2018MAR & 2018-03-29 19:00:51 & 1001 &  640 &  475.0 &   86.9 &    10.2 $\pm$    \phantom{0}4.6 &   22.7 $\pm$    \phantom{0}4.4 &   24.9 $\pm$    \phantom{0}4.5 &   32.9 $\pm$    \phantom{0}5.2 \\
&                 & 2018JUL & 2018-07-12 10:54:07 & 1100 &  640 &  475.3 &   86.1 &     1.7 $\pm$    \phantom{0}4.6 &   17.0 $\pm$    \phantom{0}4.5 &   17.1 $\pm$    \phantom{0}4.6 &   42.1 $\pm$    \phantom{0}7.8 \\
177474 & F8V      & 2017AUG & 2017-08-14 13:49:16 & 1663 &  960 &  471.3 &   89.2 &     3.4 $\pm$    \phantom{0}5.8 &$-$17.6 $\pm$    \phantom{0}5.9 &   17.9 $\pm$    \phantom{0}5.8 &  140.5 $\pm$    \phantom{0}9.9 \\
181240 & A8III    & 2018MAR & 2018-04-05 18:40:37 & 1716 & 1280 &  465.3 &   83.0 & $-$88.4 $\pm$    \phantom{0}6.9 &$-$47.5 $\pm$    \phantom{0}6.9 &  100.4 $\pm$    \phantom{0}6.9 &  104.1 $\pm$    \phantom{0}2.0 \\
&                 & 2018MAR & 2018-04-06 18:25:17 & 1726 & 1280 &  465.4 &   83.0 &$-$104.2 $\pm$    \phantom{0}6.7 &$-$54.2 $\pm$    \phantom{0}6.6 &  117.5 $\pm$    \phantom{0}6.7 &  103.7 $\pm$    \phantom{0}1.6 \\
191408 & K2.5V    & 2016JUN & 2016-06-25 12:21:58 & 3029 & 1680 &  477.2 &   90.9 &  $-$8.6 $\pm$    \phantom{0}9.3 &$-$32.0 $\pm$    \phantom{0}8.7 &   33.1 $\pm$    \phantom{0}9.0 &  127.5 $\pm$    \phantom{0}8.0 \\
192310 & K2+V     & 2016JUN & 2016-06-25 15:41:28 & 3464 & 2560 &  476.1 &   90.7 &  $-$9.8 $\pm$    \phantom{0}7.2 & $-$5.9 $\pm$    \phantom{0}7.1 &   11.4 $\pm$    \phantom{0}7.2 &  105.5 $\pm$   22.3 \\
&                 & 2018JUL & 2018-07-20 13:21:57 & 2203 & 1600 &  475.3 &   86.0 & $-$17.7 $\pm$    \phantom{0}6.9 & $-$3.5 $\pm$    \phantom{0}8.3 &   18.0 $\pm$    \phantom{0}7.6 &   95.6 $\pm$   13.9 \\
\multicolumn{1}{l}{$^R$}&& 2018JUL & 2018-07-23 11:07:49 & 1772 & 1280 &  627.0 &   84.4 & $-$14.8 $\pm$    \phantom{0}8.9 &$-$18.4 $\pm$    \phantom{0}9.8 &   23.6 $\pm$    \phantom{0}9.3 &  115.6 $\pm$   12.7\\
\multicolumn{1}{l}{$^R$}&& 2018JUL & 2018-07-23 11:33:17 & 1051 &  640 &  627.0 &   84.4 &    14.0 $\pm$   12.0 &$-$16.7 $\pm$   12.3 &   21.8 $\pm$   12.1 &  155.0 $\pm$   19.7 \\
192947 & G9III    & 2018JUL & 2018-07-18 12:25:25 & 1068 &  640 &  474.7 &   85.8 & $-$16.1 $\pm$    \phantom{0}4.7 &   15.7 $\pm$    \phantom{0}4.7 &   22.5 $\pm$    \phantom{0}4.7 &   67.9 $\pm$    \phantom{0}6.1 \\
\hline
\end{tabular}
Notes: The same aperture as used for the $\epsilon$~Sgr observations specified in Table \ref{tab:runs} in the same run was used. Spectral types (SpT) are from SIMBAD. $^R$ Indicates SDSS $r^\prime$ filter paired with R PMT detectors; all other tabulated observations used the SDSS $g^\prime$ filter paired with the B PMT detectors.\\
\end{flushleft}
\vspace{-6 pt}
\end{table*}

In Figure \ref{fig:interstellar} we present a map of the polarization of nearby stars. The stars have been selected such that their polarization is likely to be interstellar dominated. We interrogated observations made with high precision polarimeters, reported in \citet{bailey10, cotton16a, marshall16, cotton17, cotton17b, bailey17, bott18, cotton19b, cotton20b, bailey20b, marshall20, piirola20, lewis22, marshall23}, and selected stars with spectral types between A0 and K2.5 within 35$^\circ$. From this list we removed rapidly rotating stars, those in close binary systems, magnetically active stars, systems hosting significant debris disks, and any other star known to display intrinsic polarization. To this we added a small number of interstellar controls newly observed for this work (Table \ref{tab:controls}). The observations were not all made in the same bandpass so, as we have done in past work, we assumed $\lambda_{\rm max}=470$~nm and used the Serkowski Law to scale the polarization magnitude to correspond to 450~nm -- representative of the $g^\prime$ band.

The right panel of Figure \ref{fig:interstellar} illustrates that most stars near $\epsilon$~Sgr have polarizations corresponding to $\le$ 2~ppm/pc, though the closest star, HD\,169586, is a little greater. This is in line with the trends previously reported in \citet{cotton17b}, which if extrapolated predict 60.0 $\pm$ 10.5~ppm for $\epsilon$~Sgr (grey dot).

The position angles of the interstellar controls are shown in the left panel of Figure \ref{fig:interstellar}. \citet{cotton17b} presents a formula for predicting position angle of interstellar polarization for a star based on the angular separation weighted mean of nearby stars that gives 163.1$^\circ$ $\pm$ 29.1 (grey bar). However, this was only meant to apply to stars within 30~pc of the Sun, where potential controls are sparse. The paper also found better agreement than the nominal error between stars within 10$^\circ$ of each other on the sky; the median position angle of the nearest 5 stars to $\epsilon$~Sgr is 145.6$^\circ$; it is likely that the interstellar position angle is within 15-20$^\circ$ of this.

\bibliography{eps_sgr}{}

\begin{thebibliography}{}
\expandafter\ifx\csname natexlab\endcsname\relax\def\natexlab#1{#1}\fi
\providecommand{\url}[1]{\href{#1}{#1}}
\providecommand{\dodoi}[1]{doi:~\href{http://doi.org/#1}{\nolinkurl{#1}}}
\providecommand{\doeprint}[1]{\href{http://ascl.net/#1}{\nolinkurl{http://ascl.net/#1}}}
\providecommand{\doarXiv}[1]{\href{https://arxiv.org/abs/#1}{\nolinkurl{https://arxiv.org/abs/#1}}}

\bibitem[{{Anche} {et~al.}(2023){Anche}, {Douglas}, {Milani}, {Ashcraft}, {Millar-Blanchaer}, {Debes}, {Milli}, \& {Hom}}]{anche23}
{Anche}, R.~M., {Douglas}, E., {Milani}, K., {et~al.} 2023, \pasp, 135, 125001, \dodoi{10.1088/1538-3873/ad0a72}

\bibitem[{{Baade} {et~al.}(2018){Baade}, {Pigulski}, {Rivinius}, {Carciofi}, {Panoglou}, {Ghoreyshi}, {Handler}, {Kuschnig}, {Moffat}, {Pablo}, {Popowicz}, {Wade}, {Weiss}, \& {Zwintz}}]{baade18}
{Baade}, D., {Pigulski}, A., {Rivinius}, T., {et~al.} 2018, \aap, 610, A70, \dodoi{10.1051/0004-6361/201731187}

\bibitem[{{Bailey} {et~al.}(2020{\natexlab{a}}){Bailey}, {Cotton}, {Howarth}, {Lewis}, \& {Kedziora-Chudczer}}]{bailey20b}
{Bailey}, J., {Cotton}, D.~V., {Howarth}, I.~D., {Lewis}, F., \& {Kedziora-Chudczer}, L. 2020{\natexlab{a}}, \mnras, 494, 2254, \dodoi{10.1093/mnras/staa785}

\bibitem[{{Bailey} {et~al.}(2017){Bailey}, {Cotton}, \& {Kedziora-Chudczer}}]{bailey17}
{Bailey}, J., {Cotton}, D.~V., \& {Kedziora-Chudczer}, L. 2017, \mnras, 465, 1601, \dodoi{10.1093/mnras/stw2886}

\bibitem[{{Bailey} {et~al.}(2020{\natexlab{b}}){Bailey}, {Cotton}, {Kedziora-Chudczer}, {De Horta}, \& {Maybour}}]{bailey20a}
{Bailey}, J., {Cotton}, D.~V., {Kedziora-Chudczer}, L., {De Horta}, A., \& {Maybour}, D. 2020{\natexlab{b}}, \pasa, 37, e004, \dodoi{10.1017/pasa.2019.45}

\bibitem[{{Bailey} {et~al.}(2015){Bailey}, {Kedziora-Chudczer}, {Cotton}, {Bott}, {Hough}, \& {Lucas}}]{bailey15}
{Bailey}, J., {Kedziora-Chudczer}, L., {Cotton}, D.~V., {et~al.} 2015, \mnras, 449, 3064, \dodoi{10.1093/mnras/stv519}

\bibitem[{{Bailey} {et~al.}(2010){Bailey}, {Lucas}, \& {Hough}}]{bailey10}
{Bailey}, J., {Lucas}, P.~W., \& {Hough}, J.~H. 2010, \mnras, 405, 2570, \dodoi{10.1111/j.1365-2966.2010.16634.x}

\bibitem[{{Baschek} \& {Slettebak}(1988)}]{baschek88}
{Baschek}, B., \& {Slettebak}, A. 1988, \aap, 207, 112

\bibitem[{{Berghoefer} {et~al.}(1996){Berghoefer}, {Schmitt}, \& {Cassinelli}}]{berghoefer96}
{Berghoefer}, T.~W., {Schmitt}, J.~H.~M.~M., \& {Cassinelli}, J.~P. 1996, \aaps, 118, 481

\bibitem[{{Bott} {et~al.}(2018){Bott}, {Bailey}, {Cotton}, {Kedziora-Chudczer}, {Marshall}, \& {Meadows}}]{bott18}
{Bott}, K., {Bailey}, J., {Cotton}, D.~V., {et~al.} 2018, \aj, 156, 293, \dodoi{10.3847/1538-3881/aaed20}

\bibitem[{Bradley {et~al.}(2022)Bradley, Sipőcz, Robitaille, Tollerud, Vinícius, Deil, Barbary, Wilson, Busko, Donath, Günther, Cara, Lim, Meßlinger, Conseil, Bostroem, Droettboom, Bray, Bratholm, Barentsen, Craig, Rathi, Pascual, Perren, Georgiev, de~Val-Borro, Kerzendorf, Bach, Quint, \& Souchereau}]{bradley22}
Bradley, L., Sipőcz, B., Robitaille, T., {et~al.} 2022, astropy/photutils: 1.5.0, 1.5.0,  Zenodo, \dodoi{10.5281/zenodo.6825092}

\bibitem[{{Castelli} \& {Kurucz}(2003)}]{castelli03}
{Castelli}, F., \& {Kurucz}, R.~L. 2003, in IAU Symposium, Vol. 210, Modelling of Stellar Atmospheres, ed. N.~{Piskunov}, W.~W. {Weiss}, \& D.~F. {Gray}, A20.
\newblock \doarXiv{astro-ph/0405087}

\bibitem[{{Che} {et~al.}(2011){Che}, {Monnier}, {Zhao}, {Pedretti}, {Thureau}, {M{\'e}rand}, {ten Brummelaar}, {McAlister}, {Ridgway}, {Turner}, {Sturmann}, \& {Sturmann}}]{che11}
{Che}, X., {Monnier}, J.~D., {Zhao}, M., {et~al.} 2011, \apj, 732, 68, \dodoi{10.1088/0004-637X/732/2/68}

\bibitem[{{Chen} {et~al.}(2014){Chen}, {Mittal}, {Kuchner}, {Forrest}, {Lisse}, {Manoj}, {Sargent}, \& {Watson}}]{chen14}
{Chen}, C.~H., {Mittal}, T., {Kuchner}, M., {et~al.} 2014, \apjs, 211, 25, \dodoi{10.1088/0067-0049/211/2/25}

\bibitem[{{Collins} \& {Cranmer}(1991)}]{collins91b}
{Collins}, G.~W., \& {Cranmer}, S.~R. 1991, \mnras, 253, 167, \dodoi{10.1093/mnras/253.1.167}

\bibitem[{{Collins} {et~al.}(1991){Collins}, {Truax}, \& {Cranmer}}]{collins91a}
{Collins}, G.~W., {Truax}, R.~J., \& {Cranmer}, S.~R. 1991, \apjs, 77, 541, \dodoi{10.1086/191616}

\bibitem[{{Cote}(1987)}]{cote87}
{Cote}, J. 1987, \aap, 181, 77

\bibitem[{{Cotton} {et~al.}(2017{\natexlab{a}}){Cotton}, {Bailey}, {Howarth}, {Bott}, {Kedziora-Chudczer}, {Lucas}, \& {Hough}}]{cotton17}
{Cotton}, D.~V., {Bailey}, J., {Howarth}, I.~D., {et~al.} 2017{\natexlab{a}}, Nature Astronomy, 1, 690, \dodoi{10.1038/s41550-017-0238-6}

\bibitem[{{Cotton} {et~al.}(2016{\natexlab{a}}){Cotton}, {Bailey}, {Kedziora-Chudczer}, {Bott}, {Lucas}, {Hough}, \& {Marshall}}]{cotton16a}
{Cotton}, D.~V., {Bailey}, J., {Kedziora-Chudczer}, L., {et~al.} 2016{\natexlab{a}}, \mnras, 455, 1607, \dodoi{10.1093/mnras/stv2185}

\bibitem[{{Cotton} {et~al.}(2020){Cotton}, {Bailey}, {Pringle}, {Sparks}, {von Hippel}, \& {Marshall}}]{cotton20b}
{Cotton}, D.~V., {Bailey}, J., {Pringle}, J.~E., {et~al.} 2020, \mnras, 494, 4591, \dodoi{10.1093/mnras/staa1023}

\bibitem[{{Cotton} {et~al.}(2017{\natexlab{b}}){Cotton}, {Marshall}, {Bailey}, {Kedziora-Chudczer}, {Bott}, {Marsden}, \& {Carter}}]{cotton17b}
{Cotton}, D.~V., {Marshall}, J.~P., {Bailey}, J., {et~al.} 2017{\natexlab{b}}, \mnras, 467, 873, \dodoi{10.1093/mnras/stx068}

\bibitem[{{Cotton} {et~al.}(2016{\natexlab{b}}){Cotton}, {Marshall}, {Bott}, {Kedziora-Chudczer}, \& {Bailey}}]{cotton16c}
{Cotton}, D.~V., {Marshall}, J.~P., {Bott}, K., {Kedziora-Chudczer}, L., \& {Bailey}, J. 2016{\natexlab{b}}, arXiv e-prints, arXiv:1605.07742, \dodoi{10.48550/arXiv.1605.07742}

\bibitem[{{Cotton} {et~al.}(2019){Cotton}, {Marshall}, {Frisch}, {Kedziora-Chudzer}, {Bailey}, {Bott}, {Wright}, {Wyatt}, \& {Kennedy}}]{cotton19b}
{Cotton}, D.~V., {Marshall}, J.~P., {Frisch}, P.~C., {et~al.} 2019, \mnras, 483, 3636, \dodoi{10.1093/mnras/sty3318}

\bibitem[{{Cotton} {et~al.}(in prep.){Cotton}, {Bailey}, {Bott}, {Kedziora-Chudczer}, {De Horta}, {Melville}, {Filcek}, {Marshall}, {Buzasi}, {Boiko}, {Borsato}, {Perkins}, {Opitz}, {Melrose}, {Gr\"uning}, {Evensberget}, \& {Zhao}}]{cotton24}
{Cotton}, D.~V., {Bailey}, J., {Bott}, K., {et~al.} in prep., for MNRAS

\bibitem[{{Coyne}(1976)}]{coyne76}
{Coyne}, G.~V. 1976, in Be and Shell Stars, ed. A.~{Slettebak}, Vol.~70, 233

\bibitem[{{Cranmer}(2005)}]{cramner05}
{Cranmer}, S.~R. 2005, \apj, 634, 585, \dodoi{10.1086/491696}

\bibitem[{{Dekker} {et~al.}(2000){Dekker}, {D'Odorico}, {Kaufer}, {Delabre}, \& {Kotzlowski}}]{dekker00}
{Dekker}, H., {D'Odorico}, S., {Kaufer}, A., {Delabre}, B., \& {Kotzlowski}, H. 2000, in Society of Photo-Optical Instrumentation Engineers (SPIE) Conference Series, Vol. 4008, Optical and IR Telescope Instrumentation and Detectors, ed. M.~{Iye} \& A.~F. {Moorwood}, 534--545, \dodoi{10.1117/12.395512}

\bibitem[{{Domiciano de Souza} {et~al.}(2014){Domiciano de Souza}, {Kervella}, {Moser Faes}, {Dalla Vedova}, {M{\'e}rand}, {Le Bouquin}, {Espinosa Lara}, {Rieutord}, {Bendjoya}, {Carciofi}, {Hadjara}, {Millour}, \& {Vakili}}]{desouza14}
{Domiciano de Souza}, A., {Kervella}, P., {Moser Faes}, D., {et~al.} 2014, \aap, 569, A10, \dodoi{10.1051/0004-6361/201424144}

\bibitem[{{Donati}(2003)}]{donati03}
{Donati}, J.~F. 2003, in Astronomical Society of the Pacific Conference Series, Vol. 307, Solar Polarization, ed. J.~{Trujillo-Bueno} \& J.~{Sanchez Almeida}, 41

\bibitem[{{Espinosa Lara} \& {Rieutord}(2011)}]{espinosa11}
{Espinosa Lara}, F., \& {Rieutord}, M. 2011, \aap, 533, A43, \dodoi{10.1051/0004-6361/201117252}

\bibitem[{{Fitzpatrick} {et~al.}(2019){Fitzpatrick}, {Massa}, {Gordon}, {Bohlin}, \& {Clayton}}]{fitzp19}
{Fitzpatrick}, E.~L., {Massa}, D., {Gordon}, K.~D., {Bohlin}, R., \& {Clayton}, G.~C. 2019, \apj, 886, 108, \dodoi{10.3847/1538-4357/ab4c3a}

\bibitem[{{Georgy} {et~al.}(2013){Georgy}, {Ekstr{\"o}m}, {Granada}, {Meynet}, {Mowlavi}, {Eggenberger}, \& {Maeder}}]{georgy13}
{Georgy}, C., {Ekstr{\"o}m}, S., {Granada}, A., {et~al.} 2013, \aap, 553, A24, \dodoi{10.1051/0004-6361/201220558}

\bibitem[{{Golimowski} {et~al.}(1993){Golimowski}, {Durrance}, \& {Clampin}}]{golimowski93}
{Golimowski}, D.~A., {Durrance}, S.~T., \& {Clampin}, M. 1993, \aj, 105, 1108, \dodoi{10.1086/116498}

\bibitem[{{Graham} {et~al.}(2007){Graham}, {Kalas}, \& {Matthews}}]{graham07}
{Graham}, J.~R., {Kalas}, P.~G., \& {Matthews}, B.~C. 2007, \apj, 654, 595, \dodoi{10.1086/509318}

\bibitem[{{Gray} \& {Garrison}(1987)}]{gray87}
{Gray}, R.~O., \& {Garrison}, R.~F. 1987, \apjs, 65, 581, \dodoi{10.1086/191237}

\bibitem[{{Halonen} \& {Jones}(2013)}]{halonen13}
{Halonen}, R.~J., \& {Jones}, C.~E. 2013, \apj, 765, 17, \dodoi{10.1088/0004-637X/765/1/17}

\bibitem[{{Harrington} \& {Collins}(1968)}]{harrington68}
{Harrington}, J.~P., \& {Collins}, G.~W. 1968, \apj, 151, 1051, \dodoi{10.1086/149504}

\bibitem[{{Houk}(1982)}]{houk82}
{Houk}, N. 1982, {Michigan Catalogue of Two-dimensional Spectral Types for the HD stars. Volume\_3. Declinations -40 to -26} (Dept.\ of Astronomy, University of Michigan)

\bibitem[{{Howarth}(2007)}]{Howarth07}
{Howarth}, I.~D. 2007, in Astronomical Society of the Pacific Conference Series, Vol. 361, Active OB-Stars: Laboratories for Stellare and Circumstellar Physics, ed. A.~T. {Okazaki}, S.~P. {Owocki}, \& S.~{Stefl}, 15

\bibitem[{{Howarth} {et~al.}(2023){Howarth}, {Bailey}, {Cotton}, \& {Kedziora-Chudczer}}]{howarth23}
{Howarth}, I.~D., {Bailey}, J., {Cotton}, D.~V., \& {Kedziora-Chudczer}, L. 2023, \mnras, 520, 1193, \dodoi{10.1093/mnras/stad149}

\bibitem[{{Huang} {et~al.}(2010){Huang}, {Gies}, \& {McSwain}}]{Huang10}
{Huang}, W., {Gies}, D.~R., \& {McSwain}, M.~V. 2010, \apj, 722, 605, \dodoi{10.1088/0004-637X/722/1/605}

\bibitem[{{Hubeny}(2012)}]{hubeny12}
{Hubeny}, I. 2012, in IAU Symposium, Vol. 282, From Interacting Binaries to Exoplanets: Essential Modeling Tools, ed. M.~T. {Richards} \& I.~{Hubeny}, 221--228, \dodoi{10.1017/S1743921311027414}

\bibitem[{{Hubeny} {et~al.}(1985){Hubeny}, {Stefl}, \& {Harmanec}}]{hubeny85}
{Hubeny}, I., {Stefl}, S., \& {Harmanec}, P. 1985, Bulletin of the Astronomical Institutes of Czechoslovakia, 36, 214

\bibitem[{{Hubrig} {et~al.}(2001){Hubrig}, {Le Mignant}, {North}, \& {Krautter}}]{hubrig01}
{Hubrig}, S., {Le Mignant}, D., {North}, P., \& {Krautter}, J. 2001, \aap, 372, 152, \dodoi{10.1051/0004-6361:20010452}

\bibitem[{{Johnson} {et~al.}(1966){Johnson}, {Mitchell}, {Iriarte}, \& {Wisniewski}}]{johnson66}
{Johnson}, H.~L., {Mitchell}, R.~I., {Iriarte}, B., \& {Wisniewski}, W.~Z. 1966, Communications of the Lunar and Planetary Laboratory, 4, 99

\bibitem[{{Joint Iras Science}(1994)}]{iras_psc94}
{Joint Iras Science}, W.~G. 1994, {VizieR Online Data Catalog: IRAS catalogue of Point Sources, Version 2.0 (IPAC 1986)}, VizieR On-line Data Catalog: II/125. Originally published in: 1988IRASP.C......0J

\bibitem[{{Jones} {et~al.}(2022){Jones}, {Labadie-Bartz}, {Cotton}, {Naz{\'e}}, {Peters}, {Hillier}, {Neiner}, {Richardson}, {Hoffman}, {Carciofi}, {Wisniewski}, {Gayley}, {Suffak}, {Ignace}, \& {Scowen}}]{jones22}
{Jones}, C.~E., {Labadie-Bartz}, J., {Cotton}, D.~V., {et~al.} 2022, \apss, 367, 124, \dodoi{10.1007/s10509-022-04127-5}

\bibitem[{{Kastner} \& {Mazzali}(1989)}]{kastner89}
{Kastner}, J.~H., \& {Mazzali}, P.~A. 1989, \aap, 210, 295

\bibitem[{{Krivova} {et~al.}(2000){Krivova}, {Krivov}, \& {Mann}}]{krivova00}
{Krivova}, N.~A., {Krivov}, A.~V., \& {Mann}, I. 2000, \apj, 539, 424, \dodoi{10.1086/309214}

\bibitem[{{Labadie-Bartz} {et~al.}(2022){Labadie-Bartz}, {Carciofi}, {Henrique de Amorim}, {Rubio}, {Luiz Figueiredo}, {Ticiani dos Santos}, \& {Thomson-Paressant}}]{labadie-bartz22}
{Labadie-Bartz}, J., {Carciofi}, A.~C., {Henrique de Amorim}, T., {et~al.} 2022, \aj, 163, 226, \dodoi{10.3847/1538-3881/ac5abd}

\bibitem[{{Lallement} {et~al.}(2022){Lallement}, {Vergely}, {Babusiaux}, \& {Cox}}]{lallement22}
{Lallement}, R., {Vergely}, J.~L., {Babusiaux}, C., \& {Cox}, N.~L.~J. 2022, \aap, 661, A147, \dodoi{10.1051/0004-6361/202142846}

\bibitem[{{Lewis} {et~al.}(2022){Lewis}, {Bailey}, {Cotton}, {Howarth}, {Kedziora-Chudczer}, \& {van Leeuwen}}]{lewis22}
{Lewis}, F., {Bailey}, J., {Cotton}, D.~V., {et~al.} 2022, \mnras, 513, 1129, \dodoi{10.1093/mnras/stac991}

\bibitem[{{Marshall} {et~al.}(2023){Marshall}, {Cotton}, {Bott}, {Bailey}, {Kedziora-Chudczer}, \& {Brown}}]{marshall23}
{Marshall}, J.~P., {Cotton}, D.~V., {Bott}, K., {et~al.} 2023, \mnras, 522, 2777, \dodoi{10.1093/mnras/stad979}

\bibitem[{{Marshall} {et~al.}(2020){Marshall}, {Cotton}, {Scicluna}, {Bailey}, {Kedziora-Chudczer}, \& {Bott}}]{marshall20}
{Marshall}, J.~P., {Cotton}, D.~V., {Scicluna}, P., {et~al.} 2020, \mnras, 499, 5915, \dodoi{10.1093/mnras/staa3195}

\bibitem[{{Marshall} {et~al.}(2016){Marshall}, {Cotton}, {Bott}, {Ertel}, {Kennedy}, {Wyatt}, {del Burgo}, {Absil}, {Bailey}, \& {Kedziora-Chudczer}}]{marshall16}
{Marshall}, J.~P., {Cotton}, D.~V., {Bott}, K., {et~al.} 2016, \apj, 825, 124, \dodoi{10.3847/0004-637X/825/2/124}

\bibitem[{{Massa} \& {Fitzpatrick}(2000)}]{massa20}
{Massa}, D., \& {Fitzpatrick}, E.~L. 2000, \apjs, 126, 517, \dodoi{10.1086/313298}

\bibitem[{{Monnier} {et~al.}(2007){Monnier}, {Zhao}, {Pedretti}, {Thureau}, {Ireland}, {Muirhead}, {Berger}, {Millan-Gabet}, {Van Belle}, {ten Brummelaar}, {McAlister}, {Ridgway}, {Turner}, {Sturmann}, {Sturmann}, \& {Berger}}]{monnier07}
{Monnier}, J.~D., {Zhao}, M., {Pedretti}, E., {et~al.} 2007, Science, 317, 342, \dodoi{10.1126/science.1143205}

\bibitem[{{{\"O}hman}(1946)}]{ohman46}
{{\"O}hman}, Y. 1946, \apj, 104, 460, \dodoi{10.1086/144879}

\bibitem[{{Owocki}(2005)}]{owocki05}
{Owocki}, S. 2005, in Astronomical Society of the Pacific Conference Series, Vol. 337, The Nature and Evolution of Disks Around Hot Stars, ed. R.~{Ignace} \& K.~G. {Gayley}, 101

\bibitem[{{Paunzen} {et~al.}(2001){Paunzen}, {Duffee}, {Heiter}, {Kuschnig}, \& {Weiss}}]{paunzen01}
{Paunzen}, E., {Duffee}, B., {Heiter}, U., {Kuschnig}, R., \& {Weiss}, W.~W. 2001, \aap, 373, 625, \dodoi{10.1051/0004-6361:20010630}

\bibitem[{{Piirola} {et~al.}(2020){Piirola}, {Berdyugin}, {Frisch}, {Kagitani}, {Sakanoi}, {Berdyugina}, {Cole}, {Harlingten}, \& {Hill}}]{piirola20}
{Piirola}, V., {Berdyugin}, A., {Frisch}, P.~C., {et~al.} 2020, \aap, 635, A46, \dodoi{10.1051/0004-6361/201937324}

\bibitem[{{Poeckert} {et~al.}(1979){Poeckert}, {Bastien}, \& {Landstreet}}]{poeckert79}
{Poeckert}, R., {Bastien}, P., \& {Landstreet}, J.~D. 1979, \aj, 84, 812, \dodoi{10.1086/112484}

\bibitem[{{Pols} {et~al.}(1991){Pols}, {Cote}, {Waters}, \& {Heise}}]{pols91}
{Pols}, O.~R., {Cote}, J., {Waters}, L.~B.~F.~M., \& {Heise}, J. 1991, \aap, 241, 419

\bibitem[{Press {et~al.}(2007)Press, Teukolsky, Vetterling, \& Flannery}]{Numerical_recipes}
Press, W.~H., Teukolsky, S.~A., Vetterling, W.~T., \& Flannery, B.~P. 2007, Numerical Recipes 3rd Edition: The Art of Scientific Computing, 3rd edn. (New York, NY, USA: Cambridge University Press)

\bibitem[{{Quirrenbach} {et~al.}(1997){Quirrenbach}, {Bjorkman}, {Bjorkman}, {Hummel}, {Buscher}, {Armstrong}, {Mozurkewich}, {Elias}, \& {Babler}}]{quirrenbach97}
{Quirrenbach}, A., {Bjorkman}, K.~S., {Bjorkman}, J.~E., {et~al.} 1997, \apj, 479, 477, \dodoi{10.1086/303854}

\bibitem[{{Rhee} {et~al.}(2007){Rhee}, {Song}, {Zuckerman}, \& {McElwain}}]{rhee07}
{Rhee}, J.~H., {Song}, I., {Zuckerman}, B., \& {McElwain}, M. 2007, \apj, 660, 1556, \dodoi{10.1086/509912}

\bibitem[{{Ricker} {et~al.}(2015){Ricker}, {Winn}, {Vanderspek}, {Latham}, {Bakos}, {Bean}, {Berta-Thompson}, {Brown}, {Buchhave}, {Butler}, {Butler}, {Chaplin}, {Charbonneau}, {Christensen-Dalsgaard}, {Clampin}, {Deming}, {Doty}, {De Lee}, {Dressing}, {Dunham}, {Endl}, {Fressin}, {Ge}, {Henning}, {Holman}, {Howard}, {Ida}, {Jenkins}, {Jernigan}, {Johnson}, {Kaltenegger}, {Kawai}, {Kjeldsen}, {Laughlin}, {Levine}, {Lin}, {Lissauer}, {MacQueen}, {Marcy}, {McCullough}, {Morton}, {Narita}, {Paegert}, {Palle}, {Pepe}, {Pepper}, {Quirrenbach}, {Rinehart}, {Sasselov}, {Sato}, {Seager}, {Sozzetti}, {Stassun}, {Sullivan}, {Szentgyorgyi}, {Torres}, {Udry}, \& {Villasenor}}]{ricker15}
{Ricker}, G.~R., {Winn}, J.~N., {Vanderspek}, R., {et~al.} 2015, Journal of Astronomical Telescopes, Instruments, and Systems, 1, 014003, \dodoi{10.1117/1.JATIS.1.1.014003}

\bibitem[{{Rivinius} {et~al.}(2013){Rivinius}, {Carciofi}, \& {Martayan}}]{rivinius13}
{Rivinius}, T., {Carciofi}, A.~C., \& {Martayan}, C. 2013, \aapr, 21, 69, \dodoi{10.1007/s00159-013-0069-0}

\bibitem[{{Rodriguez} \& {Zuckerman}(2012)}]{rodriguez12}
{Rodriguez}, D.~R., \& {Zuckerman}, B. 2012, \apj, 745, 147, \dodoi{10.1088/0004-637X/745/2/147}

\bibitem[{{Royer} {et~al.}(2002){Royer}, {Gerbaldi}, {Faraggiana}, \& {G{\'o}mez}}]{royer02a}
{Royer}, F., {Gerbaldi}, M., {Faraggiana}, R., \& {G{\'o}mez}, A.~E. 2002, \aap, 381, 105, \dodoi{10.1051/0004-6361:20011422}

\bibitem[{{Scowen} {et~al.}(2022){Scowen}, {Gayley}, {Ignace}, {Neiner}, {Vasudevan}, {Woodruff}, {Casini}, {Shultz}, {Andersson}, \& {Wisniewski}}]{scowen22}
{Scowen}, P.~A., {Gayley}, K., {Ignace}, R., {et~al.} 2022, \apss, 367, 121, \dodoi{10.1007/s10509-022-04107-9}

\bibitem[{{Seaton}(1979)}]{seaton79}
{Seaton}, M.~J. 1979, \mnras, 187, 73, \dodoi{10.1093/mnras/187.1.73P}

\bibitem[{{Semaan} {et~al.}(2018){Semaan}, {Hubert}, {Zorec}, {Guti{\'e}rrez-Soto}, {Fr{\'e}mat}, {Martayan}, {Fabregat}, \& {Eggenberger}}]{semann18}
{Semaan}, T., {Hubert}, A.~M., {Zorec}, J., {et~al.} 2018, \aap, 613, A70, \dodoi{10.1051/0004-6361/201629243}

\bibitem[{{Serkowski}(1973)}]{Serkowski1973}
{Serkowski}, K. 1973, in IAU Symposium, Vol.~52, Interstellar Dust and Related Topics, ed. J.~M. {Greenberg} \& H.~C. {van de Hulst}, 145

\bibitem[{{Serkowski} {et~al.}(1975){Serkowski}, {Mathewson}, \& {Ford}}]{serkowski75}
{Serkowski}, K., {Mathewson}, D.~S., \& {Ford}, V.~L. 1975, \apj, 196, 261, \dodoi{10.1086/153410}

\bibitem[{{Shepard} {et~al.}(2022){Shepard}, {Gies}, {Kaper}, \& {De Koter}}]{shepard22}
{Shepard}, K., {Gies}, D.~R., {Kaper}, L., \& {De Koter}, A. 2022, \apj, 931, 35, \dodoi{10.3847/1538-4357/ac66e6}

\bibitem[{{Slettebak}(1975)}]{slettback75b}
{Slettebak}, A. 1975, \apj, 197, 137, \dodoi{10.1086/153493}

\bibitem[{{Spurr}(2006)}]{spurr06}
{Spurr}, R. J.~D. 2006, \jqsrt, 102, 316, \dodoi{10.1016/j.jqsrt.2006.05.005}

\bibitem[{{Stetson}(1987)}]{stetson87}
{Stetson}, P.~B. 1987, \pasp, 99, 191, \dodoi{10.1086/131977}

\bibitem[{{Tinney} {et~al.}(2004){Tinney}, {Ryder}, {Ellis}, {Churilov}, {Dawson}, {Smith}, {Waller}, {Whittard}, {Haynes}, {Lankshear}, {Barton}, {Evans}, {Shortridge}, {Farrell}, \& {Bailey}}]{tinney04}
{Tinney}, C.~G., {Ryder}, S.~D., {Ellis}, S.~C., {et~al.} 2004, in Society of Photo-Optical Instrumentation Engineers (SPIE) Conference Series, Vol. 5492, Ground-based Instrumentation for Astronomy, ed. A.~F.~M. {Moorwood} \& M.~{Iye}, 998--1009, \dodoi{10.1117/12.550980}

\bibitem[{{Townsend} {et~al.}(2004){Townsend}, {Owocki}, \& {Howarth}}]{townsend04}
{Townsend}, R.~H.~D., {Owocki}, S.~P., \& {Howarth}, I.~D. 2004, \mnras, 350, 189, \dodoi{10.1111/j.1365-2966.2004.07627.x}

\bibitem[{\VAN{Belle}{van}{van}~Belle(2012)}]{vanbelle12}
\VAN{Belle}{van}{van}~Belle, G.~T. 2012, \aapr, 20, 51, \dodoi{10.1007/s00159-012-0051-2}

\bibitem[{{Vandeportal} {et~al.}(2019){Vandeportal}, {Bastien}, {Simon}, {Augereau}, \& {Storer}}]{vandeportal19}
{Vandeportal}, J., {Bastien}, P., {Simon}, A., {Augereau}, J.-C., \& {Storer}, {\'E}. 2019, \mnras, 483, 3510, \dodoi{10.1093/mnras/sty3060}

\bibitem[{\VAN{Leeuwen}{van}{van}~Leeuwen(2007)}]{vanLeeuwen07}
\VAN{Leeuwen}{van}{van}~Leeuwen, F. 2007, \aap, 474, 653, \dodoi{10.1051/0004-6361:20078357}

\bibitem[{{Walker} {et~al.}(2005){Walker}, {Kuschnig}, {Matthews}, {Cameron}, {Saio}, {Lee}, {Kambe}, {Masuda}, {Guenther}, {Moffat}, {Rucinski}, {Sasselov}, \& {Weiss}}]{walker05}
{Walker}, G.~A.~H., {Kuschnig}, R., {Matthews}, J.~M., {et~al.} 2005, \apjl, 635, L77, \dodoi{10.1086/499362}

\bibitem[{{Whittet} {et~al.}(1992){Whittet}, {Martin}, {Hough}, {Rouse}, {Bailey}, \& {Axon}}]{whittet92}
{Whittet}, D.~C.~B., {Martin}, P.~G., {Hough}, J.~H., {et~al.} 1992, \apj, 386, 562, \dodoi{10.1086/171039}

\bibitem[{{Wilking} {et~al.}(1982){Wilking}, {Lebofsky}, \& {Rieke}}]{wilking82}
{Wilking}, B.~A., {Lebofsky}, M.~J., \& {Rieke}, G.~H. 1982, \aj, 87, 695, \dodoi{10.1086/113147}

\bibitem[{{Wood} {et~al.}(1997){Wood}, {Bjorkman}, \& {Bjorkman}}]{wood97}
{Wood}, K., {Bjorkman}, K.~S., \& {Bjorkman}, J.~E. 1997, \apj, 477, 926, \dodoi{10.1086/303747}

\bibitem[{{Zhao} {et~al.}(2009){Zhao}, {Monnier}, {Pedretti}, {Thureau}, {M{\'e}rand}, {ten Brummelaar}, {McAlister}, {Ridgway}, {Turner}, {Sturmann}, {Sturmann}, {Goldfinger}, \& {Farrington}}]{zhao09}
{Zhao}, M., {Monnier}, J.~D., {Pedretti}, E., {et~al.} 2009, \apj, 701, 209, \dodoi{10.1088/0004-637X/701/1/209}

\bibitem[{{Zorec} {et~al.}(2016){Zorec}, {Fr{\'e}mat}, {Domiciano de Souza}, {Royer}, {Cidale}, {Hubert}, {Semaan}, {Martayan}, {Cochetti}, {Arias}, {Aidelman}, \& {Stee}}]{zorec16}
{Zorec}, J., {Fr{\'e}mat}, Y., {Domiciano de Souza}, A., {et~al.} 2016, \aap, 595, A132, \dodoi{10.1051/0004-6361/201628760}

\end{thebibliography}
\bibliographystyle{aasjournal}



\end{document}